\pdfoutput=1
\documentclass[12pt,a4paper]{article}

\usepackage{ifthen} 
\newboolean{pdflatex}
\setboolean{pdflatex}{true} 

\newboolean{articletitles}
\setboolean{articletitles}{true} 

\newboolean{uprightparticles}
\setboolean{uprightparticles}{false} 

\def\paperauthors{LHCb collaboration} 
\def\paperasciititle{Search for the lepton-flavour violating decays B0->K*0mue and Bs->phimue} 
\def\papertitle{Search for the lepton-flavour violating decays $\decay{\Bd}{\Kstarz\mu^\pm e^\mp}$ and $\decay{\Bs}{\phi\mu^\pm e^\mp}$} 
\def\paperkeywords{{High Energy Physics}, {LHCb}} 
\def\papercopyright{\the\year\ CERN for the benefit of the LHCb collaboration} 
\def\paperlicence{CC BY 4.0 licence}
\def\paperlicenceurl{https://creativecommons.org/licenses/by/4.0/}

 \def\cls       {\ensuremath{{\rm CL}_{\rm s}}\xspace}

\usepackage[top=1in, bottom=1.25in, left=1in, right=1in]{geometry}

\usepackage{microtype}
\usepackage{lineno}  
\usepackage{xspace} 
\usepackage{caption} 

\usepackage{graphicx}  
\usepackage{color}
\usepackage{colortbl}
\graphicspath{{./figs/}} 

\usepackage{amsmath} 
\usepackage{amssymb}
\usepackage{amsfonts}
\usepackage{upgreek} 

\newcommand*\patchAmsMathEnvironmentForLineno[1]{%
\expandafter\let\csname old#1\expandafter\endcsname\csname #1\endcsname
\expandafter\let\csname oldend#1\expandafter\endcsname\csname
end#1\endcsname
 \renewenvironment{#1}%
   {\linenomath\csname old#1\endcsname}%
   {\csname oldend#1\endcsname\endlinenomath}%
}
\newcommand*\patchBothAmsMathEnvironmentsForLineno[1]{%
  \patchAmsMathEnvironmentForLineno{#1}%
  \patchAmsMathEnvironmentForLineno{#1*}%
}
\AtBeginDocument{%
\patchBothAmsMathEnvironmentsForLineno{equation}%
\patchBothAmsMathEnvironmentsForLineno{align}%
\patchBothAmsMathEnvironmentsForLineno{flalign}%
\patchBothAmsMathEnvironmentsForLineno{alignat}%
\patchBothAmsMathEnvironmentsForLineno{gather}%
\patchBothAmsMathEnvironmentsForLineno{multline}%
\patchBothAmsMathEnvironmentsForLineno{eqnarray}%
}

\usepackage{hyperxmp}

\usepackage[pdftex,
            pdfauthor={\paperauthors},
            pdftitle={\paperasciititle},
            pdfkeywords={\paperkeywords},
            pdfcopyright={Copyright (C) \papercopyright},
            pdflicenseurl={\paperlicenceurl}]{hyperref}

\usepackage[colorinlistoftodos,textsize=scriptsize]{todonotes}

\usepackage[bottom,flushmargin,hang,multiple]{footmisc}

\usepackage[all]{hypcap} 

\usepackage{xspace} 
\usepackage{upgreek}

\def\lhcb   {\mbox{LHCb}\xspace}

\def\MagUp {\mbox{\em Mag\kern -0.05em Up}\xspace}

\ifthenelse{\boolean{uprightparticles}}%
{

 \def\Pmu         {\ensuremath{\upmu}\xspace}

 \def\Ppi         {\ensuremath{\uppi}\xspace}

 \def\Pphi        {\ensuremath{\upphi}\xspace}

 \def\Ppsi        {\ensuremath{\uppsi}\xspace}

 \def\PDelta      {\ensuremath{\Delta}\xspace}                 
 \def\PXi         {\ensuremath{\Xi}\xspace}                 
 \def\PLambda     {\ensuremath{\Lambda}\xspace}                 
 \def\PSigma      {\ensuremath{\Sigma}\xspace}                 
 \def\POmega      {\ensuremath{\Omega}\xspace}                 
 \def\PUpsilon    {\ensuremath{\Upsilon}\xspace}
 \let\oldPi\Pi
 \def\PPi         {\ensuremath{\oldPi}\xspace}

 \def\PB      {\ensuremath{\mathrm{B}}\xspace}                 
                  
 \def\PD      {\ensuremath{\mathrm{D}}\xspace}

 \def\PJ      {\ensuremath{\mathrm{J}}\xspace}                 
 \def\PK      {\ensuremath{\mathrm{K}}\xspace}

 \def\Pb      {\ensuremath{\mathrm{b}}\xspace}                 
 \def\Pc      {\ensuremath{\mathrm{c}}\xspace}                 
 \def\Pd      {\ensuremath{\mathrm{d}}\xspace}                 
 \def\Pe      {\ensuremath{\mathrm{e}}\xspace}

 \def\Pi      {\ensuremath{\mathrm{i}}\xspace}

 \def\Pp      {\ensuremath{\mathrm{p}}\xspace}

 \def\Ps      {\ensuremath{\mathrm{s}}\xspace}

 \def\thebaroffset{0.0em}
}
{

 \def\Pmu         {\ensuremath{\mu}\xspace}

 \def\Ppi         {\ensuremath{\pi}\xspace}

 \def\Pphi        {\ensuremath{\phi}\xspace}

 \def\Ppsi        {\ensuremath{\psi}\xspace}                 
                  
 \mathchardef\PDelta="7101
 \mathchardef\PXi="7104
 \mathchardef\PLambda="7103
 \mathchardef\PSigma="7106
 \mathchardef\POmega="710A
 \mathchardef\PUpsilon="7107
 \mathchardef\PPi="7105
                  
 \def\PB      {\ensuremath{B}\xspace}                 
                  
 \def\PD      {\ensuremath{D}\xspace}

 \def\PJ      {\ensuremath{J}\xspace}                 
 \def\PK      {\ensuremath{K}\xspace}

 \def\Pb      {\ensuremath{b}\xspace}                 
 \def\Pc      {\ensuremath{c}\xspace}                 
 \def\Pd      {\ensuremath{d}\xspace}                 
 \def\Pe      {\ensuremath{e}\xspace}

 \def\Pi      {\ensuremath{i}\xspace}

 \def\Pp      {\ensuremath{p}\xspace}

 \def\Ps      {\ensuremath{s}\xspace}

 \def\thebaroffset{0.18em}
}
\newcommand{\offsetoverline}[2][\thebaroffset]{\kern #1\overline{\kern -#1 #2}}%

\makeatletter
\ifcase \@ptsize \relax
  \newcommand{\miniscule}{\@setfontsize\miniscule{4}{5}}
\or
  \newcommand{\miniscule}{\@setfontsize\miniscule{5}{6}}
\or
  \newcommand{\miniscule}{\@setfontsize\miniscule{5}{6}}
\fi
\makeatother

\DeclareRobustCommand{\optbar}[1]{\shortstack{{\miniscule (\rule[.5ex]{1.25em}{.18mm})}
  \\ [-.7ex] $#1$}}

\def\en         {{\ensuremath{\Pe^-}}\xspace}   
\def\ep         {{\ensuremath{\Pe^+}}\xspace}
 
\def\emp        {{\ensuremath{\Pe^\mp}}\xspace} 
\def\epem       {{\ensuremath{\Pe^+\Pe^-}}\xspace}

\def\mup        {{\ensuremath{\Pmu^+}}\xspace}
\def\mun        {{\ensuremath{\Pmu^-}}\xspace} 
\def\mupm       {{\ensuremath{\Pmu^\pm}}\xspace} 
 
\def\mumu       {{\ensuremath{\Pmu^+\Pmu^-}}\xspace}

\def\ellell     {\ensuremath{\ell^+ \ell^-}\xspace}

\def\dquark    {{\ensuremath{\Pd}}\xspace}
\def\dquarkbar {{\ensuremath{\overline \dquark}}\xspace}

\def\squark    {{\ensuremath{\Ps}}\xspace}
\def\squarkbar {{\ensuremath{\overline \squark}}\xspace}

\def\cquark    {{\ensuremath{\Pc}}\xspace}

\def\bquark    {{\ensuremath{\Pb}}\xspace}
\def\bquarkbar {{\ensuremath{\overline \bquark}}\xspace}

\def\pion   {{\ensuremath{\Ppi}}\xspace}

\def\pim    {{\ensuremath{\pion^-}}\xspace}

\def\kaon    {{\ensuremath{\PK}}\xspace}
\def\Kbar    {{\ensuremath{\offsetoverline{\PK}}}\xspace}

\def\KorKbar {\kern \thebaroffset\optbar{\kern -\thebaroffset \PK}{}\xspace}

\def\Kp      {{\ensuremath{\kaon^+}}\xspace}
\def\Km      {{\ensuremath{\kaon^-}}\xspace}

\def\Kstarz  {{\ensuremath{\kaon^{*0}}}\xspace}
\def\Kstarzb {{\ensuremath{\Kbar{}^{*0}}}\xspace}

\newcommand{\phiz}{\ensuremath{\Pphi}\xspace}

\def\Dbar    {{\ensuremath{\offsetoverline{\PD}}}\xspace}
\def\D       {{\ensuremath{\PD}}\xspace}

\def\DorDbar {\kern \thebaroffset\optbar{\kern -\thebaroffset \PD}\xspace}

\def\Dzb     {{\ensuremath{\Dbar{}^0}}\xspace}
\def\Dp      {{\ensuremath{\D^+}}\xspace}
\def\Dm      {{\ensuremath{\D^-}}\xspace}

\def\DpDm    {\ensuremath{\Dp {\kern -0.16em \Dm}}\xspace}

\def\B       {{\ensuremath{\PB}}\xspace}

\def\BorBbar {\kern \thebaroffset\optbar{\kern -\thebaroffset \PB}\xspace}

\def\Bd      {{\ensuremath{\B^0}}\xspace}

\def\BdorBdbar {\kern \thebaroffset\optbar{\kern -\thebaroffset \Bd}\xspace}
\def\Bu      {{\ensuremath{\B^+}}\xspace}

\def\Bs      {{\ensuremath{\B^0_\squark}}\xspace}

\def\BsorBsbar {\kern \thebaroffset\optbar{\kern -\thebaroffset \Bs}\xspace}

\def\jpsi     {{\ensuremath{{\PJ\mskip -3mu/\mskip -2mu\Ppsi}}}\xspace}
\def\psitwos  {{\ensuremath{\Ppsi{(2S)}}}\xspace}

\def\Y#1S{\ensuremath{\PUpsilon{(#1S)}}\xspace}

\def\proton      {{\ensuremath{\Pp}}\xspace}

\def\Lz          {{\ensuremath{\PLambda}}\xspace}

\def\LorLbar     {\kern \thebaroffset\optbar{\kern -\thebaroffset \PLambda}\xspace}

\def\Lb           {{\ensuremath{\Lz^0_\bquark}}\xspace}

\def\BF         {{\ensuremath{\mathcal{B}}}\xspace}

\newcommand{\decay}[2]{\ensuremath{#1\!\to #2}\xspace} 

\def\to                 {\ensuremath{\rightarrow}\xspace}

\def\qsq       {{\ensuremath{q^2}}\xspace}

\def\AT#1     {\ensuremath{A_{\mathrm{T}}^{#1}}\xspace}           

\def\ctl       {\ensuremath{\cos{\theta_\ell}}\xspace}
\def\ctk       {\ensuremath{\cos{\theta_K}}\xspace}

\def\C#1      {\ensuremath{\mathcal{C}_{#1}}\xspace}                       
\def\Cp#1     {\ensuremath{\mathcal{C}_{#1}^{'}}\xspace}                    
\def\Ceff#1   {\ensuremath{\mathcal{C}_{#1}^{\mathrm{(eff)}}}\xspace}        
\def\Cpeff#1  {\ensuremath{\mathcal{C}_{#1}^{'\mathrm{(eff)}}}\xspace}       
\def\Ope#1    {\ensuremath{\mathcal{O}_{#1}}\xspace}                       
\def\Opep#1   {\ensuremath{\mathcal{O}_{#1}^{'}}\xspace}                    

\newcommand{\nospaceunit}[1]{\ensuremath{\text{#1}}}       
\newcommand{\aunit}[1]{\ensuremath{\text{\,#1}}}       

\newcommand{\tev}{\aunit{Te\kern -0.1em V}\xspace}
\newcommand{\gev}{\aunit{Ge\kern -0.1em V}\xspace}
\newcommand{\mev}{\aunit{Me\kern -0.1em V}\xspace}
\newcommand{\kev}{\aunit{ke\kern -0.1em V}\xspace}
\newcommand{\ev}{\aunit{e\kern -0.1em V}\xspace}
 
\newcommand{\mevc}{\ensuremath{\aunit{Me\kern -0.1em V\!/}c}\xspace}
\newcommand{\gevc}{\ensuremath{\aunit{Ge\kern -0.1em V\!/}c}\xspace}
\newcommand{\mevcc}{\ensuremath{\aunit{Me\kern -0.1em V\!/}c^2}\xspace}
\newcommand{\gevcc}{\ensuremath{\aunit{Ge\kern -0.1em V\!/}c^2}\xspace}

\def\mum  {\ensuremath{\,\upmu\nospaceunit{m}}\xspace}

\def\fb   {\ensuremath{\aunit{fb}}\xspace}
\def\invfb   {\ensuremath{\fb^{-1}}\xspace}

\def\gsim{{~\raise.15em\hbox{$>$}\kern-.85em
          \lower.35em\hbox{$\sim$}~}\xspace}
\def\lsim{{~\raise.15em\hbox{$<$}\kern-.85em
          \lower.35em\hbox{$\sim$}~}\xspace}

\def\pt         {\ensuremath{p_{\mathrm{T}}}\xspace}

\def\ptot       {\ensuremath{p}\xspace}

\def\evtgen     {\mbox{\textsc{EvtGen}}\xspace}

\def\geant      {\mbox{\textsc{Geant4}}\xspace}

\def\photos     {\mbox{\textsc{Photos}}\xspace}

\def\pythia     {\mbox{\textsc{Pythia}}\xspace}

\def\tell1  {TELL1\xspace}
\def\ukl1   {UKL1\xspace}

\newcommand{\lhcborcid}[1]{\href{https://orcid.org/#1}{\hspace*{0.1em}\raisebox{-0.45ex}{\includegraphics[width=1em]{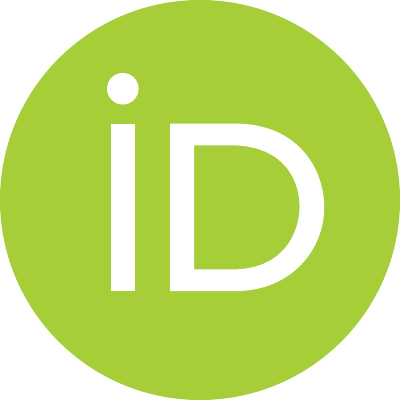}}}}

\usepackage{cite} 
\usepackage{mciteplus}

\usepackage{longtable} 

\begin{document}

\renewcommand{\thefootnote}{\fnsymbol{footnote}}
\setcounter{footnote}{1}
\begin{titlepage}
\pagenumbering{roman}

\vspace*{-1.5cm}
\centerline{\large EUROPEAN ORGANIZATION FOR NUCLEAR RESEARCH (CERN)}
\vspace*{1.5cm}
\noindent
\begin{tabular*}{\linewidth}{lc@{\extracolsep{\fill}}r@{\extracolsep{0pt}}}
\ifthenelse{\boolean{pdflatex}}
{\vspace*{-1.5cm}\mbox{\!\!\!\includegraphics[width=.14\textwidth]{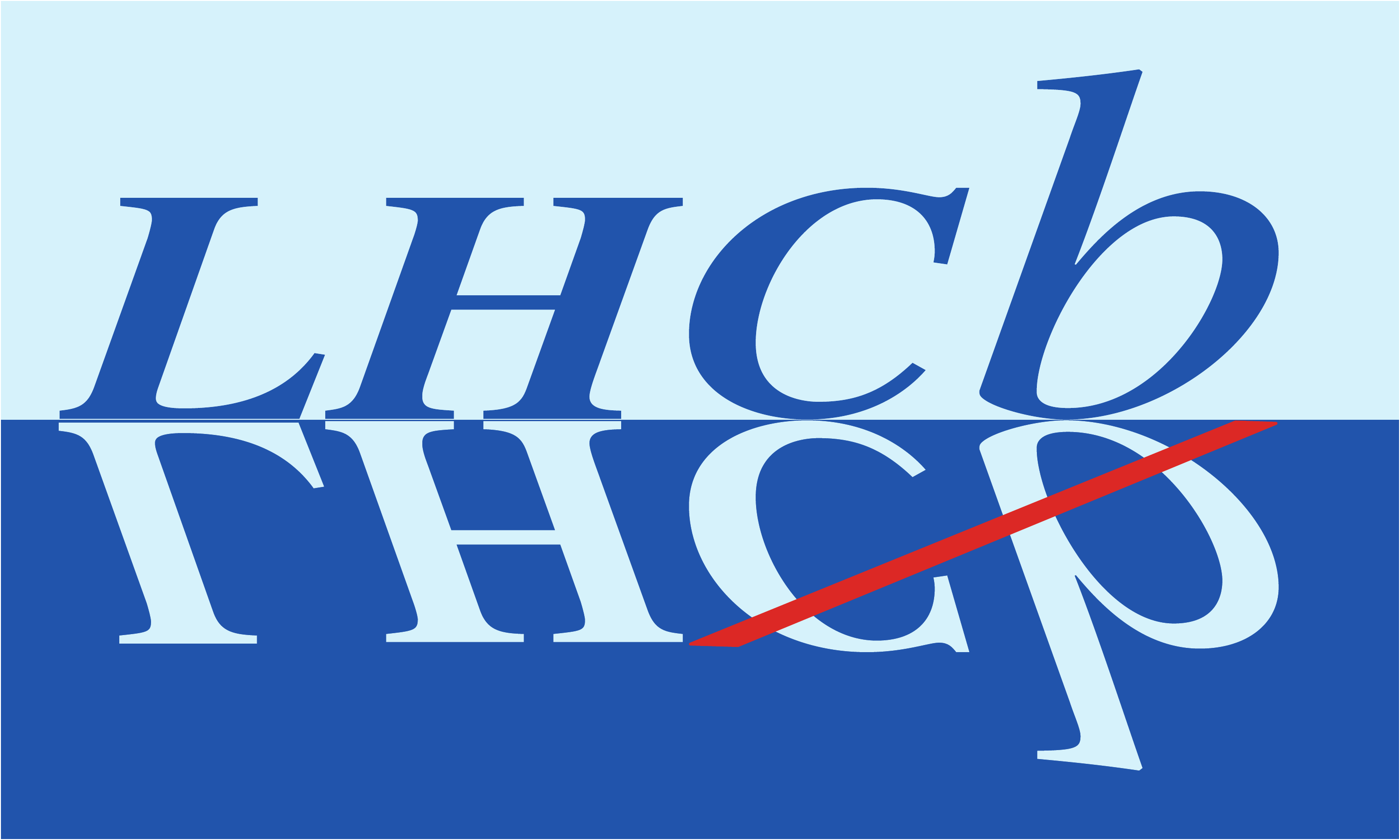}} & &}%
{\vspace*{-1.2cm}\mbox{\!\!\!\includegraphics[width=.12\textwidth]{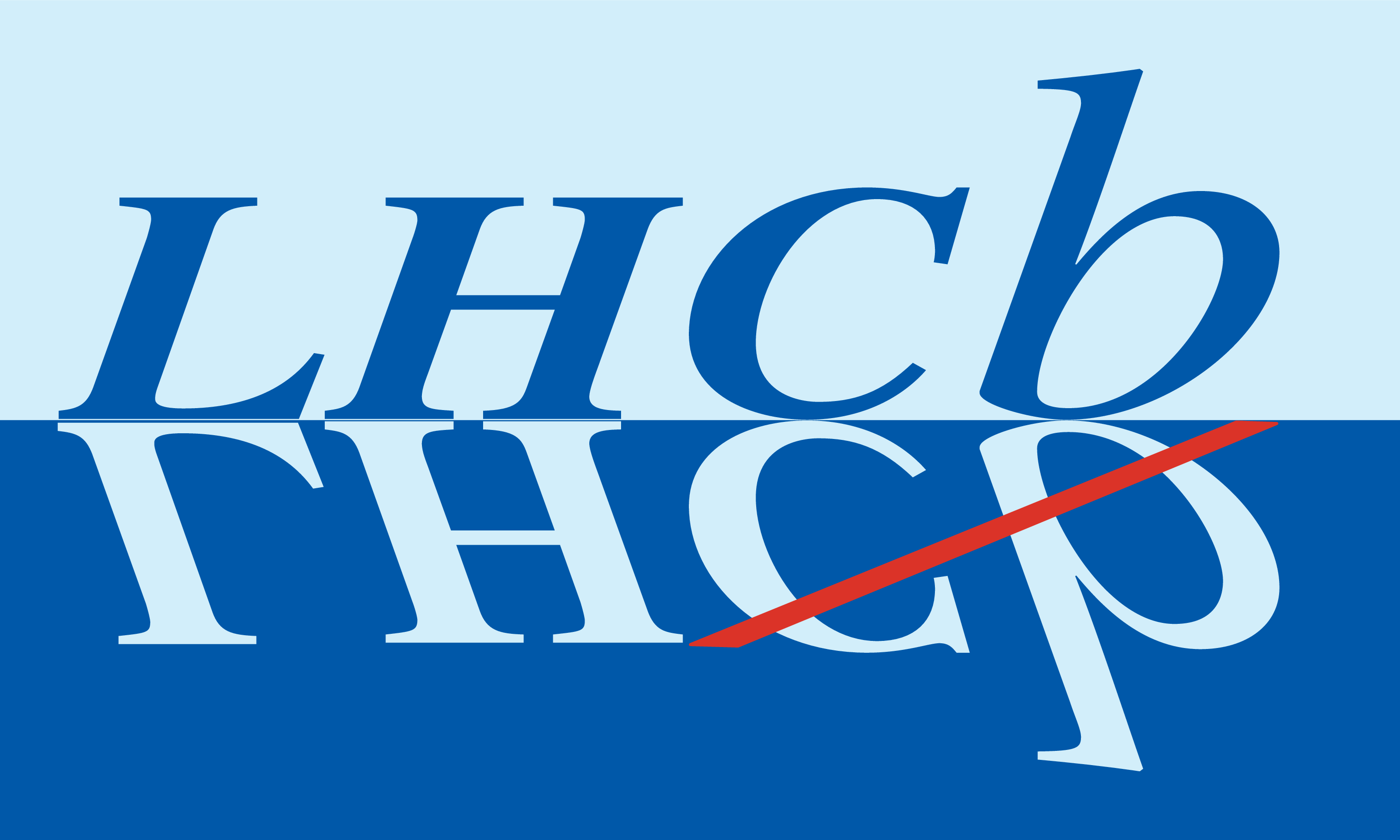}} & &}%
\\
 & & CERN-EP-2022-097 \\  
 & & LHCb-PAPER-2022-008 \\  
 & & June 15, 2023 \\
 & & \\
\end{tabular*}

\vspace*{4.0cm}

{\normalfont\bfseries\boldmath\huge
\begin{center}
  \papertitle 
\end{center}
}

\vspace*{1.5cm}

\begin{center}
\paperauthors\footnote{Authors are listed at the end of this paper.}
\end{center}

\vspace{\fill}

\begin{abstract}
  \noindent
  A search for the lepton-flavour violating decays $\decay{\Bd}{\Kstarz\mu^\pm e^\mp}$ and $\decay{\Bs}{\phi\mu^\pm e^\mp}$ is presented,
  using proton-proton collision data collected by the LHCb detector at the LHC, corresponding to an integrated luminosity of $9\invfb$.
  No significant signals are observed and upper limits of
  \begin{align*}
    {\cal B}(\decay{\Bd}{\Kstarz\mu^+ e^-}) &< \phantom{1}5.7\times 10^{-9}~(6.9\times 10^{-9}),\\
    {\cal B}(\decay{\Bd}{\Kstarz\mu^- e^+}) &< \phantom{1}6.8\times 10^{-9}~(7.9\times 10^{-9}),\\
    {\cal B}(\decay{\Bd}{\Kstarz\mu^\pm e^\mp}) &< 10.1\times 10^{-9}~(11.7\times 10^{-9}),\\
    {\cal B}(\decay{\Bs}{\phi\mu^\pm e^\mp}) &< 16.0\times 10^{-9}~(19.8\times 10^{-9})
  \end{align*}
  are set at 90\%~(95\%) confidence level. 
  These results constitute the world's most stringent limits to date, 
  with the limit on the decay $\decay{\Bs}{\phi\mu^\pm e^\mp}$ the first being set.
  In addition, limits are reported for scalar and left-handed lepton-flavour violating New Physics scenarios. 
\end{abstract}

\vspace*{1.5cm}

\begin{center}
  Published in JHEP 06 (2023) 073
\end{center}

\vspace{\fill}

{\footnotesize 
\centerline{\copyright~\papercopyright. \href{\paperlicenceurl}{\paperlicence}.}}
\vspace*{2mm}

\end{titlepage}


\newpage
\setcounter{page}{2}
\mbox{~}

\renewcommand{\thefootnote}{\arabic{footnote}}
\setcounter{footnote}{0}

\cleardoublepage


\pagestyle{plain} 
\setcounter{page}{1}
\pagenumbering{arabic}


\section{Introduction}
\label{sec:introduction}
Processes that are forbidden or strongly suppressed in the Standard Model~(SM) are sensitive to new heavy particles beyond the SM, 
and can probe energy scales beyond those accessible with direct searches. 
Lepton-flavour violating (LFV) decays are forbidden in the SM,
but the observation of neutrino oscillations shows the existence of LFV in the neutral lepton sector. 
An observation of LFV decays involving charged leptons would constitute a clear and unambiguous sign of New Physics (NP). 

Recently, 
studies of rare decays of $b$-hadrons  
have received considerable attention due to the appearance of the flavour anomalies in rare $\decay{b}{s\ellell}$ transitions~\cite{LHCb-PAPER-2013-017,LHCb-PAPER-2014-006,LHCb-PAPER-2015-009,LHCb-PAPER-2015-023,LHCb-PAPER-2016-012, LHCb-PAPER-2021-014,LHCb-PAPER-2015-051,LHCb-PAPER-2020-002,LHCb-PAPER-2020-041,Aaboud:2018krd,Khachatryan:2015isa,Sirunyan:2017dhj,LHCb-PAPER-2014-024,LHCb-PAPER-2017-013,LHCb-PAPER-2019-009,LHCb-PAPER-2019-040,LHCb-PAPER-2021-004,LHCb-PAPER-2021-038,Wehle:2016yoi,Lees:2012tva,BELLE:2019xld,Belle:2019oag}. 
The tensions with SM predictions seen in lepton flavour universality tests~\cite{LHCb-PAPER-2014-024,LHCb-PAPER-2017-013,LHCb-PAPER-2019-009,LHCb-PAPER-2019-040,LHCb-PAPER-2021-004,LHCb-PAPER-2021-038,Wehle:2016yoi,Lees:2012tva,BELLE:2019xld,Belle:2019oag}, in particular, motivate searches for LFV $b$-hadron decays, 
as lepton flavour non-universality is closely connected with LFV~\cite{Glashow:2014iga}. 
Specific NP scenarios that can induce LFV $b$-hadron decays include models with scalar or vector leptoquarks~\cite{Hiller:2016kry,deMedeirosVarzielas:2015yxm,Crivellin:2017dsk},
and models with additional $Z^\prime$ bosons~\cite{Crivellin:2015era}.
Branching fractions for $\decay{\bquarkbar}{\squarkbar \mu^\pm e^\mp}$ decays like $\decay{\Bd}{\Kstarz\mu^\pm e^\mp}$ could be as large as ${\cal O}(10^{-7})$~\cite{Crivellin:2015era},
close to the currently best limit of $\BF(\decay{\Bd}{\Kstarz\mu^\pm e^\mp}) < 1.8\times 10^{-7}$ at $90\%$ confidence level (CL) obtained by the Belle collaboration~\cite{Belle:2018peg}. 

The LHCb collaboration has performed searches for $\decay{\bquarkbar}{\squarkbar\mu^\pm e^\mp}$ and $\decay{\bquarkbar}{\dquarkbar\mu^\pm e^\mp}$ transitions 
using the decays $\decay{B^0_{(s)}}{\mu^{\pm}e^{\mp}}$~\cite{LHCb-PAPER-2017-031} and $\decay{\Bu}{\Kp\mu^{\pm}e^{\mp}}$~\cite{LHCb-PAPER-2019-022},
resulting in exclusion limits of $\BF(\decay{\Bd}{\mu^{\pm}e^{\mp}})<1.0 \times 10^{-9}$, $\BF(\decay{\Bs}{\mu^{\pm}e^{\mp}})<5.4\times 10^{-9}$, $\BF(\decay{\Bu}{\Kp\mup\en})<6.4\times 10^{-9}$, and $\BF(\decay{\Bu}{\Kp\mun\ep})<7.0\times 10^{-9}$ at $90\%$ CL. 
Both analyses were performed using the LHCb Run~1 data sample, corresponding to an integrated luminosity of $3\invfb$. 

This paper presents a search for the lepton-flavour violating decays $\decay{\Bd}{\Kstarz(\to\Kp\pim)\mu^\pm e^\mp}$ and $\decay{\Bs}{\phi(\to\Kp\Km)\mu^\pm e^\mp}$. 
The inclusion of charge-conjugate processes is implied throughout.
The symbols $\Kstarz$ and $\phi$ refer to the $\kaon^{*}(892)^{0}$ and $\phi(1020)$ vector mesons.  
The search uses the data sample collected by the LHCb experiment in proton-proton ($\proton\proton$) collisions at centre-of-mass energies of $7\tev$ ($2011$), $8\tev$ ($2012$), and $13\tev$ ($2015\text{--}2018$), 
corresponding to a total integrated luminosity of $9\invfb$. 
As NP models can affect $\decay{\bquarkbar}{\squarkbar\mup\en}$ and $\decay{\bquarkbar}{\squarkbar\mun\ep}$ transitions differently, limits for the decays $\decay{\Bd}{\Kstarz \mup\en}$ (referred to as same-sign due to muon and kaon charge being equal) and $\decay{\Bd}{\Kstarz \ep\mun}$ (referred to as opposite-sign) are also reported separately. 

The tree level decays $\decay{\Bd}{\jpsi(\to\mumu)\Kstarz}$ and $\decay{\Bs}{\jpsi(\to\mumu)\phi}$, which exhibit large yields and have a final state similar to the signal decays, are used as normalisation channels.
To avoid experimenters' bias, the candidates in the signal regions, defined using the invariant masses of the final state as $m(\Kp\pim\mu^\pm e^\mp)\in[4900,5600]\mevcc$ and $m(\Kp\Km\mu^\pm e^\mp)\in[4900,5600]\mevcc$, were not examined until the selections and limit setting procedures were finalised. 

\section{Detector and simulation}
\label{sec:detector}
The \lhcb detector~\cite{LHCb-DP-2008-001,LHCb-DP-2014-002} is a single-arm forward
spectrometer covering the \mbox{pseudorapidity} range $2<\eta <5$,
designed for the study of particles containing \bquark or \cquark
quarks. The detector includes a high-precision tracking system
consisting of a silicon-strip vertex detector surrounding the $pp$
interaction region~\cite{LHCb-DP-2014-001}, a large-area silicon-strip detector located
upstream of a dipole magnet with a bending power of about
$4{\mathrm{\,Tm}}$, and three stations of silicon-strip detectors and straw
drift tubes~\cite{LHCb-DP-2013-003,LHCb-DP-2017-001} 
placed downstream of the magnet.
The tracking system provides a measurement of the momentum, \ptot, of charged particles with
a relative uncertainty that varies from 0.5\% at low momentum to 1.0\% at 200\gevc.
The minimum distance of a track to a primary $pp$ collision vertex (PV), the impact parameter (IP), 
is measured with a resolution of $(15+29/\pt)\mum$,
where \pt is the component of the momentum transverse to the beam, in\,\gevc.
Different types of charged hadrons are distinguished using information
from two ring-imaging Cherenkov detectors~\cite{LHCb-DP-2012-003}. 
Photons, electrons and hadrons are identified by a calorimeter system consisting of
scintillating-pad and preshower detectors, an electromagnetic
and a hadronic calorimeter. Muons are identified by a
system composed of alternating layers of iron and multiwire
proportional chambers~\cite{LHCb-DP-2012-002,LHCb-DP-2013-001}.

The online event selection is performed by a trigger system~\cite{LHCb-DP-2012-004}.
Signal candidates first need to pass the hardware trigger~(L0), 
which requires the final-state muon to have sizeable \pt. 
In the subsequent software trigger, a full event reconstruction is performed. 
The software trigger requires a two-, three- or four-track
secondary vertex with a significant displacement from any primary
$pp$ interaction vertex. At least one charged particle
must have a transverse momentum $\pt > 1.6\gevc$ ($\pt>1.0\gevc$ if the particle is identified as muon) and be 
inconsistent with originating from a PV. 
A multivariate algorithm~\cite{BBDT,LHCb-PROC-2015-018} is used for
the identification of secondary vertices consistent with the decay
of a \bquark hadron.

Simulated samples are required to
determine the reconstruction and selection efficiencies,
and to estimate contributions from residual backgrounds. 
In the simulation, $pp$ collisions are generated using
\pythia~\cite{Sjostrand:2007gs,*Sjostrand:2006za} 
with a specific \lhcb configuration~\cite{LHCb-PROC-2010-056}.
Decays of unstable particles
are described by \evtgen~\cite{Lange:2001uf}, in which final-state
radiation is generated using \photos~\cite{davidson2015photos}.
The interaction of the generated particles with the detector, and its response,
are implemented using the \geant
toolkit~\cite{Allison:2006ve, *Agostinelli:2002hh} as described in
Ref.~\cite{LHCb-PROC-2011-006}. 
The underlying $pp$ interaction is reused multiple times, with an independently generated signal decay for each~\cite{LHCb-DP-2018-004}. 
Residual mismodelling of the particle identification and tracking performance,
the \pt\ spectra of \Bd\ and \Bs\ mesons,
track multiplicity, 
and the efficiency of the L0 trigger are calibrated using high-yield control samples from data. 

\section{Selection of signal candidates}
\label{sec:selection}
Candidates for $\decay{\Bd}{\Kstarz\mu^\pm e^\mp}$ and $\decay{\Bs}{\phi\mu^\pm e^\mp}$ signal decays are reconstructed in the
$\Kp\pim \mu^\pm e^\mp$ and $\Kp\Km \mu^\pm e^\mp$ final states, respectively. 
Stringent particle identification criteria are applied to the final-state hadrons and leptons, using information from the Cherenkov detectors, the muon chambers, and the calorimeter system. 
The final-state tracks are required to have significant $\chi^2_{\rm IP}$ with respect to any PV in the event, where $\chi^2_{\rm IP}$ is defined as the difference in the vertex-fit $\chi^2$ of a given PV reconstructed with and without the track being considered. 
The four final-state tracks are fit to a secondary vertex (SV), which needs to have good fit quality and be significantly displaced from any PV in the event. 
The invariant mass of the $\Kp\pim$ ($\Kp\Km$) system is required to be within $100\mevcc$ ($12\mevcc$) of the known $\Kstarz$~($\phi$) mass~\cite{pdg2020}. 
Furthermore, the reconstructed $B^0_{(s)}$ mass of signal candidates is required to be in the range $[4300,6700]\mevcc$. 

Dedicated vetoes are applied to reject backgrounds originating from misidentified $b$-hadron decays, referred to as peaking backgrounds. 
The decays $\decay{\Bd}{\jpsi(\to\ellell)\Kstarz}$ and $\decay{\Bd}{\psitwos(\to\ellell)\Kstarz}$ ($\decay{\Bs}{\jpsi(\to\ellell)\phi}$ and $\decay{\Bs}{\psitwos(\to\ellell)\phi}$) can be a source of background if one of the leptons is misidentified as the different lepton species. 
Candidates for which the reconstructed \Bd\ (\Bs) mass 
is in the range of $[5200,5350]\mevcc$ ($[5300,5450]\mevcc$) are rejected, where the reconstructed \Bd\ (\Bs) mass is determined in a fit forcing the dilepton system to have the known mass of the $\jpsi$ or $\psitwos$ meson~\cite{pdg2020}.
The same tree-level decays to charmonia can also mimic the signal decay if a lepton is misidentified as a hadron and vice-versa. 
To suppress these contributions, criteria on the invariant mass of the system comprised of a hadron and the lepton of opposite charge are applied, where the lepton is assigned the hadron mass hypothesis.
Candidates are rejected if the resulting invariant mass is in the range $[800,1050]\mevcc$ ($[1000,1040]\mevcc$) and therefore consistent with the decay of a $\Kstarz$~($\phi$) meson.
Semileptonic $\decay{b}{c(\to s\ell^{\prime +}\nu_{\ell^\prime})\ell^-\bar{\nu}_\ell}$ cascade decays can result in the same charged final-state tracks in the detector as signal decays.
These decays are rejected by a stringent requirement on  $m(\Kp\pim\ell^\pm)>2\gevcc$ ($m(\Kp\Km\ell^\pm)>2\gevcc$), 
where the invariant mass is calculated using the lepton momentum without the addition of potential bremsstrahlung photons. 
This removes contributions from $\decay{\Bd}{D^{(*)-}\ell^+\nu_\ell}$ ($\decay{\Bs}{D_s^{(*)-}\ell^+\nu_\ell}$) decays.
Semileptonic $B_{(s)}^0$ decays to higher 
excited $D_{(s)}^-$ 
resonances can escape this veto due to their higher masses; 
they are therefore modelled in the fit as discussed in Sec.~\ref{sec:fitmodel}. 
Background from several further sources are studied and found to be suppressed to negligible levels by the stringent particle identification criteria; 
these include rare $\decay{\bquarkbar}{\squarkbar\ellell}$ decays like $\decay{\Bd}{\Kstarz\ellell}$, and $\decay{\Bs}{\phi\ellell}$ (with lepton and potentially also hadron misidentification), 
as well as fully hadronic $\decay{\Bd}{\Kstarz \pi^+\pi^-}$ and  $\decay{\Bs}{\phi \pi^+\pi^-}$ decays (with misidentification of the pions as leptons). 

Background from combinations of random tracks (combinatorial background) is reduced using a boosted decision tree (BDT)~\cite{Breiman} classifier trained with the AdaBoost algorithm~\cite{AdaBoost} as implemented in the TMVA software package~\cite{Hocker:2007ht,*TMVA4}.
The BDT classifier is trained separately for the two signal decays using calibrated simulation as signal.
Data from the $B_{(s)}^0$ upper mass sideband region $[5600,6700]\mevcc$ are used as a proxy for the background.
The classifier is trained using a $k$-folding approach and its performance is verified using standard cross-validation techniques~\cite{Blum:1999:BHB:307400.307439}. 
The classifier uses the (transverse) momentum of the $B_{(s)}^0$ candidate, 
its vertex fit quality and flight distance significance,
the angle between the $B_{(s)}^0$ momentum and the vector connecting the associated PV and the $B_{(s)}^0$ decay vertex, 
and the $\chi^2_{\rm IP}$ of the $B_{(s)}^0$ candidate and the  final-state particles. 
The selection criterion on the classifier output is chosen according to the Punzi figure of merit $\varepsilon_{\rm sig}/(3/2+\sqrt{N_{\rm comb}})$~\cite{Punzi:2003bu}. 
Here, $\varepsilon_{\rm sig}$ denotes the signal efficiency and $N_{\rm comb}$ the expected combinatorial background yield, 
which is extrapolated from the upper mass sideband using the reconstructed same-sign lepton samples $\Kp\pim \mu^\pm e^\pm$ and $\Kp\Km \mu^\pm e^\pm$. 
Relative to the previously described selection criteria, the BDT requirement results in a signal efficiency of $55\text{--}80\%$, depending on the signal mode, and a rejection for combinatorial background of larger than $99\%$.

The normalisation modes, $\decay{\Bd}{\jpsi(\to\mumu)\Kstarz}$ and $\decay{\Bs}{\jpsi(\to\mumu)\phi}$, are reconstructed and selected in a way that is as similar as possible to the corresponding signal decays. 
In contrast to the signal modes, the muon identification criteria are applied to both final-state leptons. 
In addition, the invariant mass of the dimuon system is required to be within $\pm60\mevcc$ of the known \jpsi\ mass~\cite{pdg2020},
and the dedicated vetoes against $b$-hadron decays to charmonia are removed.  
For the $\decay{\Bd}{\jpsi\Kstarz}$ normalisation mode, an additional veto on the invariant mass of the $\Kp\mumu$ system rejects candidates with $m(\Kp\mumu)\in [5200,5400]\mevcc$. 
The decays $\decay{\Bd}{\jpsi\Kstarz}$ and $\decay{\Bs}{\jpsi\phi}$, with both $\decay{\jpsi}{\ep\en}$ and $\decay{\jpsi}{\mumu}$, are used as control modes to validate simulation and to model the signal mass resolution as discussed in Sec.~\ref{sec:fitmodel}. 
The control decays $\decay{\Bd}{\jpsi(\to\ep\en)\Kstarz}$ and $\decay{\Bs}{\jpsi(\ep\en)\phi}$ are selected in the $m(\ep\en)$ mass region $[2400,3300]\mevcc$.

\section{Normalisation}
\label{sec:normalisation}
The signal yields $N_{\rm sig}$ are obtained from a fit of the reconstructed $B_{(s)}^0$ mass distribution and translated to signal branching fractions $\BF_{\rm sig}$ using the normalisation modes $\decay{\Bd}{\jpsi(\to\mumu)\Kstarz}$ and $\decay{\Bs}{\jpsi(\to\mumu)\phi}$ according to
\begin{equation}
  \BF_{\rm sig} = \underbrace{\frac{\BF_{\rm norm}}{N_{\rm norm}} \times \frac{\varepsilon_{\rm norm}}{\varepsilon_{\rm sig}}}_{\textstyle = \alpha} \times N_{\rm sig}, \label{eq:bfformula}
\end{equation}
where $\BF_{\rm norm}$ denotes the branching fractions of the normalisation mode, given by $\BF(\decay{\Bd}{\jpsi(\to\mumu)\Kstarz})=(7.57\pm 0.30)\times 10^{-5}$ or $\BF(\decay{\Bs}{\jpsi(\to\mumu)\phi})=(6.07\pm 0.29)\times 10^{-5}$~\cite{pdg2020,LHCb-PAPER-2020-046}, $N_{\rm norm}$ the yields of the normalisation mode, and $\alpha$ the normalisation constant.

The efficiency ratio between normalisation and signal decays, $\varepsilon_{\rm norm}/\varepsilon_{\rm sig}$, is determined using calibrated simulation and found to be
around $2.8$ ($2.6$)
for the decays $\decay{\Bd}{\jpsi\Kstarz}$ and $\decay{\Bd}{\Kstarz\mu^\pm e^\mp}$ ($\decay{\Bs}{\jpsi\phi}$ and $\decay{\Bs}{\phi\mu^\pm e^\mp}$).
The efficiencies for the normalisation modes are significantly higher than for the signal decays as they exhibit higher trigger efficiencies, more efficient particle identification, and background vetoes with higher efficiencies.

\begin{figure}
  \centering
\includegraphics[width=0.49\textwidth]{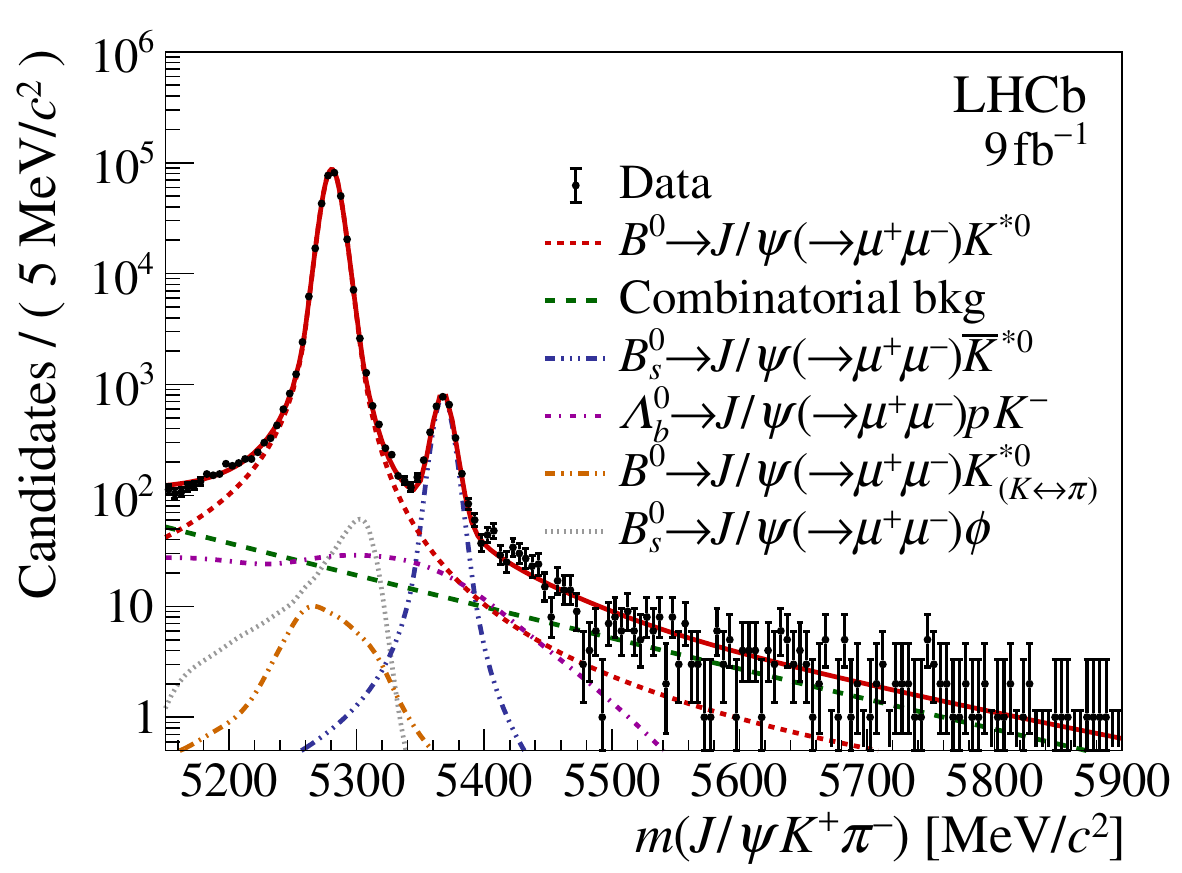}
\includegraphics[width=0.49\textwidth]{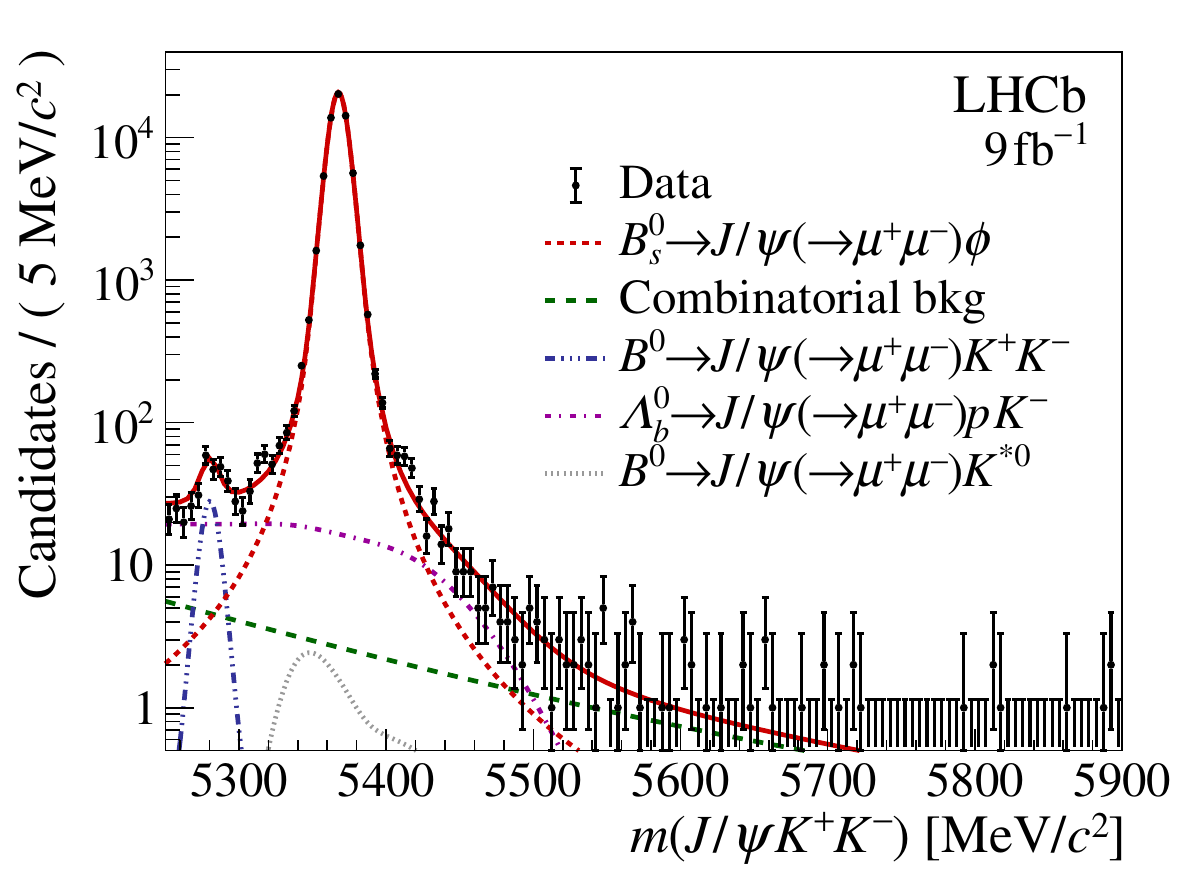}
  \caption{\small
    Mass distributions for the normalisation channels (left) $\decay{\Bd}{\jpsi(\to\mumu)\Kstarz}$ and (right) $\decay{\Bs}{\jpsi(\to\mumu)\phi}$ combining the different data taking periods, overlaid with the fit results.\label{fig:normalisation}}
\end{figure}

The yields for the normalisation modes are determined using an unbinned extended maximum-likelihood fit to the 
reconstructed $B_{(s)}^0$ mass distribution, 
in which the invariant mass of the dimuon system is constrained to the known \jpsi\ mass. 
The fit range is limited to $[5150,5900]\mevcc$ in the \Bd mode ($[5250,5900]\mevcc$ in the \Bs mode) to avoid partially reconstructed $b$-hadron decays at low invariant masses. 
The signal distribution is modelled using a double-sided Hypatia function~\cite{Santos:2013gra}. 
The shape parameters of the Hypatia function are determined using simulation, except for a resolution and mass shift parameter 
which are left to float 
in the fit to data to allow for mismodelling.  
In the fit of the decay $\decay{\Bd}{\jpsi\Kstarz}$, a component for the CKM-suppressed mode $\decay{\Bs}{\jpsi\Kstarzb}$ is also included, 
which is modelled identically to the normalisation mode, but shifted by the known $\Bs-\Bd$ mass difference~\cite{pdg2020}.
The $\decay{\Bs}{\jpsi\Kstarzb}$ component is Gaussian constrained to its expectation using simulation. 
Similarly, in the $\decay{\Bs}{\jpsi\phiz}$ normalisation mode fit, a small component of the decay $\decay{\Bd}{\jpsi\Kp\Km}$ must be accounted for, where the invariant mass of the non-resonant $\Kp\Km$ system overlaps with the $\phiz$ selection.
Again, the normalisation channel model is used to describe this component, including a mass shift by the known $\Bd-\Bs$ mass difference.
The background yield is floated in the fit.
Residual backgrounds from misidentified $b$-hadron decays to $\jpsi$ final states are modelled using kernel density estimates obtained from calibrated simulation.
For the decay $\decay{\Bd}{\jpsi\Kstarz}$, misidentified backgrounds from
$\decay{\Lb}{\jpsi p\Km}$, $\decay{\Bs}{\jpsi \phi}$, and $\decay{\Bd}{\jpsi\Kstarz}$ (with $K\leftrightarrow \pi$ misidentification), are included in the fit.
For the decay $\decay{\Bs}{\jpsi\phi}$, misidentified backgrounds from 
$\decay{\Lb}{\jpsi p\Km}$ and $\decay{\Bd}{\jpsi\Kstarz}$ decays are included. 
The yields of all backgrounds from misidentification are constrained to their expectations using Gaussian functions. 
The last component of the fit for the normalisation yields is the combinatorial background, which is modelled using an exponential function,
with its slope and normalisation allowed to vary freely. 
The $m(\jpsi\Kp\pim)$ and $m(\jpsi\Kp\Km)$ distributions, overlaid with the fit results combined over the different data taking periods, are shown in Fig.~\ref{fig:normalisation}.
The obtained normalisation channel yields are given in Tab.~\ref{tab:normalisation}. 
The resulting normalisation constants $\alpha$ for the different data taking periods are given in Tab.~\ref{tab:alpha}. 

\begin{table}
    \caption{\small
      Normalisation mode yields $[10^3]$ for different periods of data taking.\label{tab:normalisation}}
  \centering
  \begin{tabular}{lrrr}\hline
          & \multicolumn{3}{c}{Yield [$10^3$]}\\
    Mode & \multicolumn{1}{c}{$2011\text{--}2012$} & \multicolumn{1}{c}{$2015\text{--}2016$} & \multicolumn{1}{c}{$2017\text{--}2018$}\\ \hline
    $\decay{\Bd}{\decay{\jpsi(}{\mumu)}\Kstarz}$ & $88.88\pm 0.30$ & $85.56\pm 0.29$ & $139.05\pm 0.37$\\
    $\decay{\Bs}{\decay{\jpsi(}{\mumu)}\phi}$ & $17.89\pm 0.13$ & $16.21\pm 0.13$ & $30.59\pm 0.18$\\
    \hline\end{tabular}
\end{table}

\begin{table}
    \caption{\small
    Normalisation constant $\alpha$ $[10^{-9}]$ with associated statistical and systematic uncertainties, added in quadrature, for different periods of data taking.
    The total uncertainty is dominated by systematic effects, which are discussed in Sec.~\ref{sec:systematics}.
    The year-to-year $B^0/B_s^0$ ratio variation is due to different BDT criteria against combinatorial background, tuned individually for each data taking period and mode.
    \label{tab:alpha}}
    \centering
  \begin{tabular}{lrrr}\hline
          & \multicolumn{3}{c}{$\alpha \pm (\sigma_{\rm stat}\oplus\sigma_{\rm syst})$ [$10^{-9}$]}\\
    Mode & \multicolumn{1}{c}{$2011\text{--}2012$} & \multicolumn{1}{c}{$2015\text{--}2016$} & \multicolumn{1}{c}{$2017\text{--}2018$}\\ \hline

    $\decay{\Bd}{\Kstarz\mu^+ e^-}$     & $2.47\pm 0.14 $   & $2.38\pm 0.16$    & $1.49\pm 0.09$    \\
    $\decay{\Bd}{\Kstarz\mu^- e^+}$     & $2.50\pm 0.15$    & $2.39\pm 0.16$    & $1.49\pm 0.09$    \\
    $\decay{\Bd}{\Kstarz\mu^\pm e^\mp}$ & $2.48\pm 0.14$    & $2.39\pm 0.16$    & $1.49\pm 0.09$    \\
    $\decay{\Bs}{\phi\mu^\pm e^\mp}$    & $9.50\pm 0.70$    & $9.68\pm 0.78$    & $5.09\pm 0.39$    \\

    \hline\end{tabular}
\end{table}

\section{Signal fit}
\label{sec:fitmodel}
The signal decays $\decay{\Bd}{\Kstarz\mu^\pm e^\mp}$ and $\decay{\Bs}{\phi\mu^\pm e^\mp}$ are modelled using the sum of two Crystal Ball functions~\cite{Skwarnicki:1986xj},
with power-law tails on either sides.
The shape parameters are determined on simulation and fixed in the fit to data. 
Corrections to the mass resolution
are determined using the control decays
$\decay{\Bd}{\jpsi\Kstarz}$ and $\decay{\Bs}{\jpsi\phi}$. 
Information from both $\decay{\jpsi}{\mumu}$ and $\decay{\jpsi}{\epem}$ final states is combined to determine these correction factors,
as no appropriate control mode with a $\mu^\pm e^\mp$ final state exists. 
Depending on the data taking period, the correction factors scale the core Gaussian widths by $1.04\text{--}1.08$ with uncertainties at the percent level.
In the nominal fit, the uncertainties on the scale factors are included as Gaussian constraints. 

Semileptonic cascade decays involving higher excited $D_{(s)}^{-}$ resonances are modelled in the fit as they can pass the selection requirement $m(\Kp\pim\ell^\pm)>2\gevcc$ ($m(\Kp\Km\ell^\pm)>2\gevcc$) due to their high masses. 
Their shapes are modelled using kernel density estimates of fully simulated $\decay{\Bd}{D_2^{*}(2460)^- (\to\Dzb(\to\Kp\ell^-\bar{\nu}_\ell )\pim) \ell^{\prime +}\nu_{\ell^{\prime}}}$ and $\decay{\Bs}{D_{s2}^{*}(2573)^- (\to\Dzb(\to\Kp\ell^-\bar{\nu}_\ell )\Km) \ell^{\prime +}\nu_{\ell^{\prime}}}$ decays, as these decays are one of the dominant contributions of the remaining background from semileptonic cascades.
Alternative models that include potential contributions from $D_1(2420)$ and $D_0^*(2300)$ states have also been considered. 
%
For the decay $\decay{\Bd}{\Kstarz\mu^\pm e^\mp}$, an additional background contribution arises from  $\decay{\Bu}{\Dzb(\to \Kp\ell^-\bar{\nu}_\ell)\ell^{\prime +}\nu_{\ell^{\prime}}}$ decays, which are combined with a random $\pim$ from the event and thus can pass the veto against semileptonic cascade decays.  
This background is modelled using a kernel density estimate from simulated samples. 
Its yield is Gaussian constrained to the expectation from simulation.
%
The combinatorial background is modelled with a single exponential function.
Due to the low residual combinatorial background yield in the signal data samples,
the exponential slope is constrained from a fit to same-sign lepton data samples with the reconstructed final states $\Kp\pim\mu^\pm e^\pm$ and $\Kp\Km\mu^\pm e^\pm$,
using a relaxed BDT cut. 
A systematic uncertainty is assigned to this choice, as discussed in Sec.~\ref{sec:systematics}. 
The combinatorial background yields are allowed to float independently in each data taking periods. 

\begin{figure}
  \centering
  \includegraphics[width=0.49\textwidth]{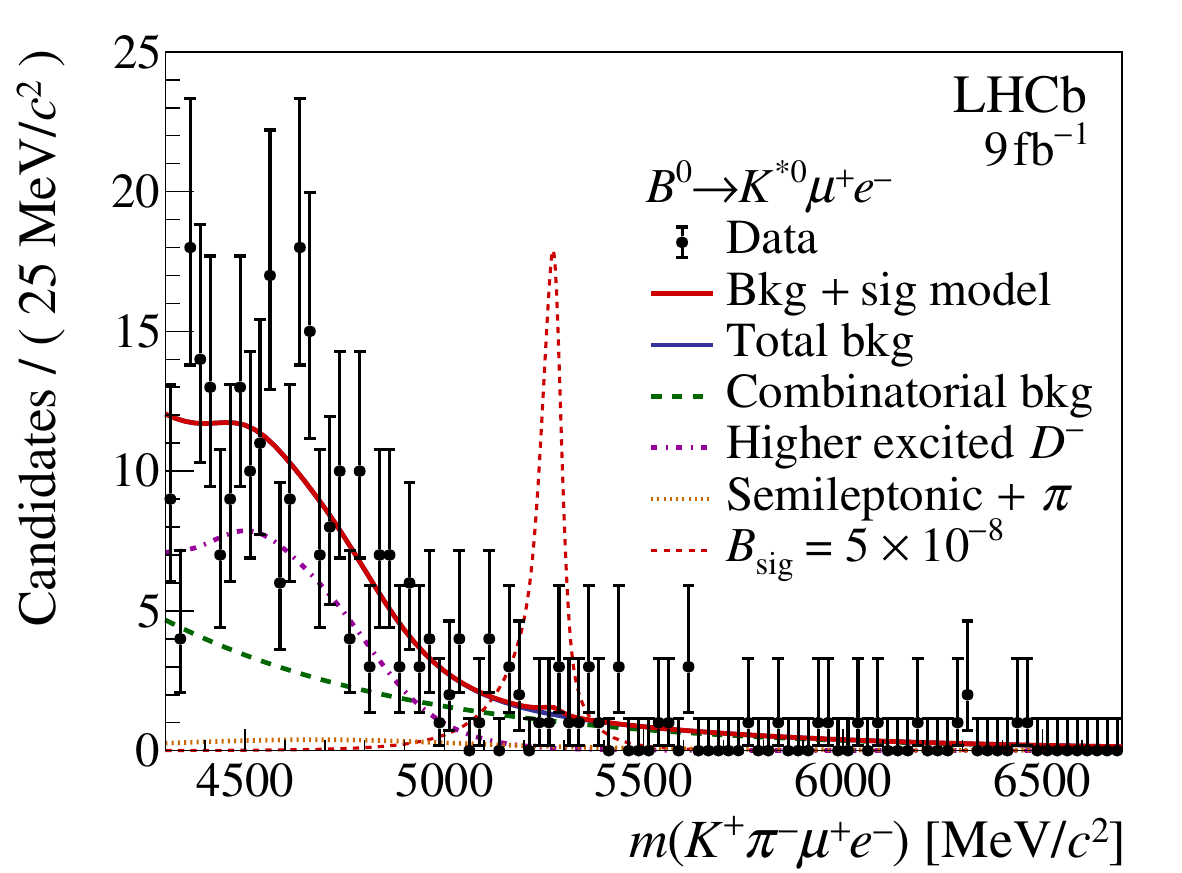}
  \includegraphics[width=0.49\textwidth]{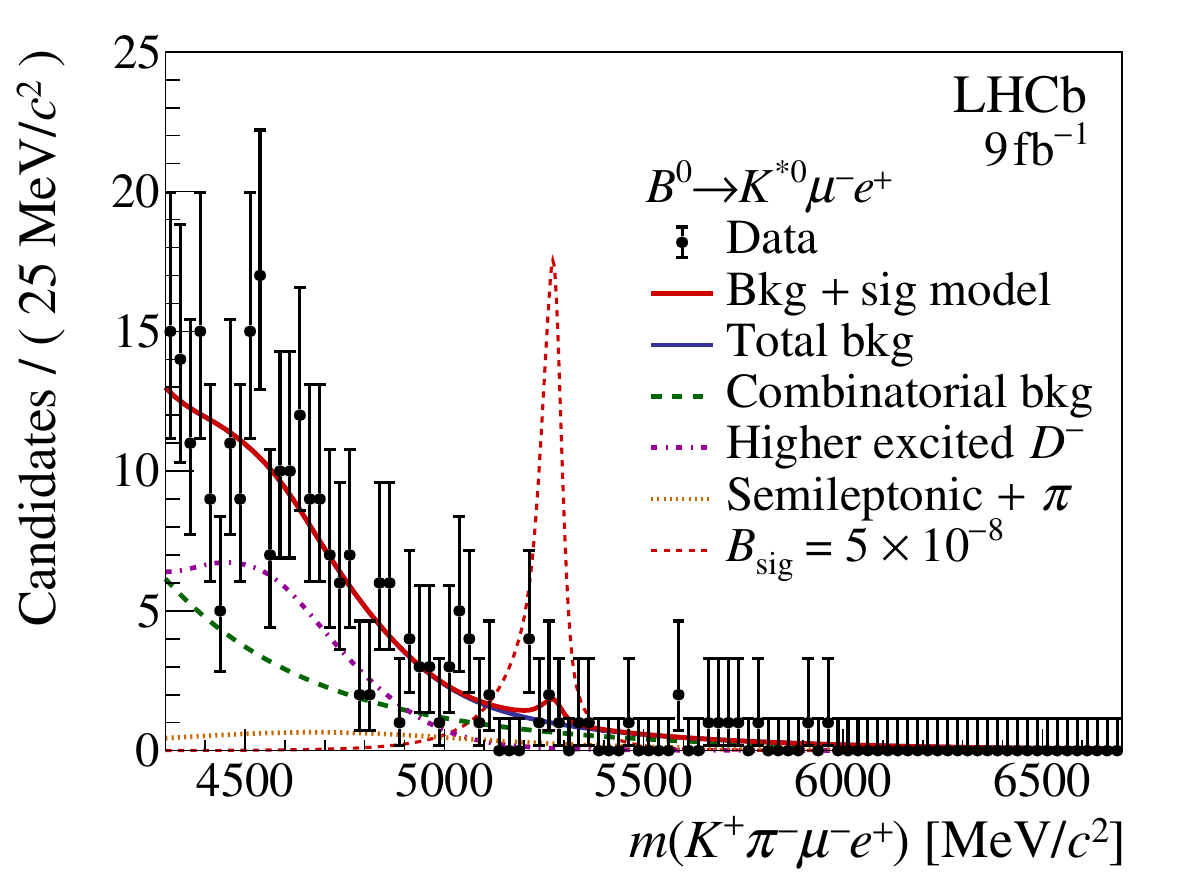}\\
    \includegraphics[width=0.49\textwidth]{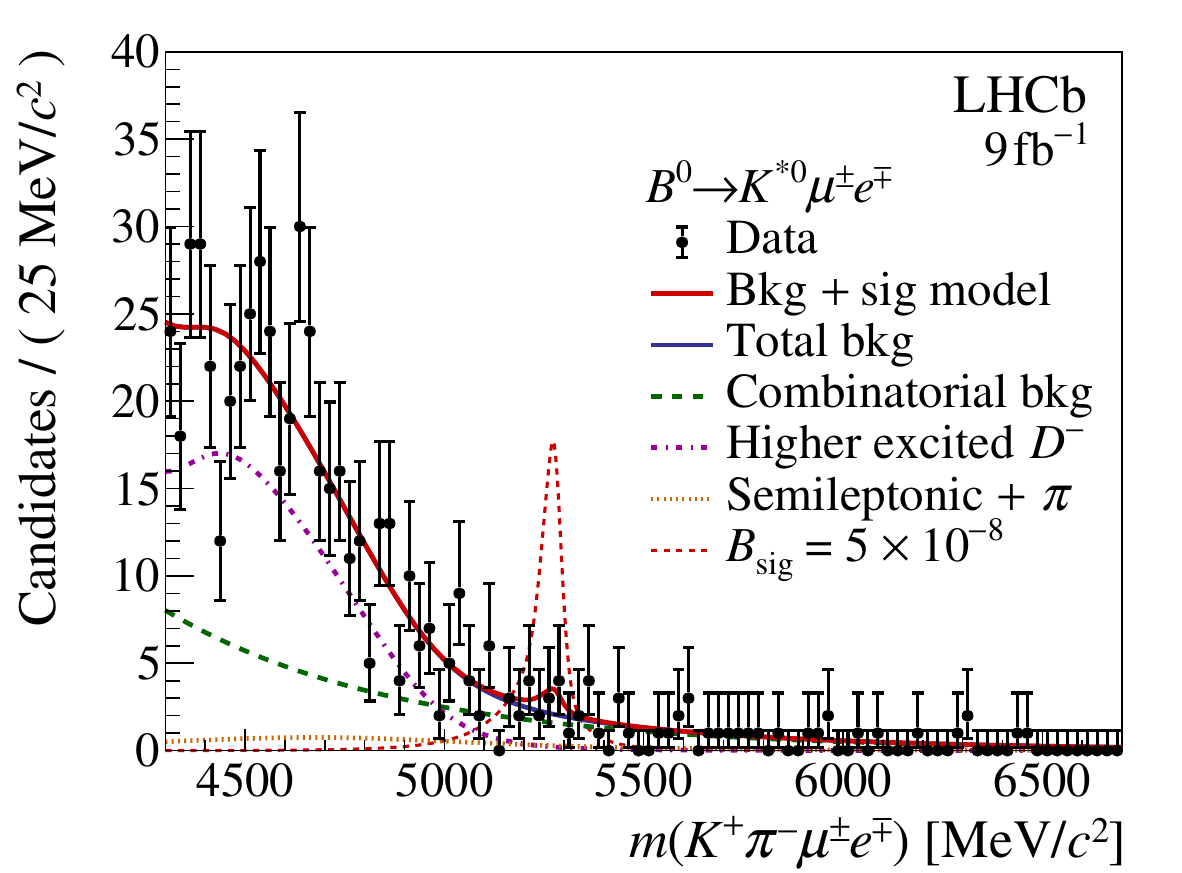}
  \includegraphics[width=0.49\textwidth]{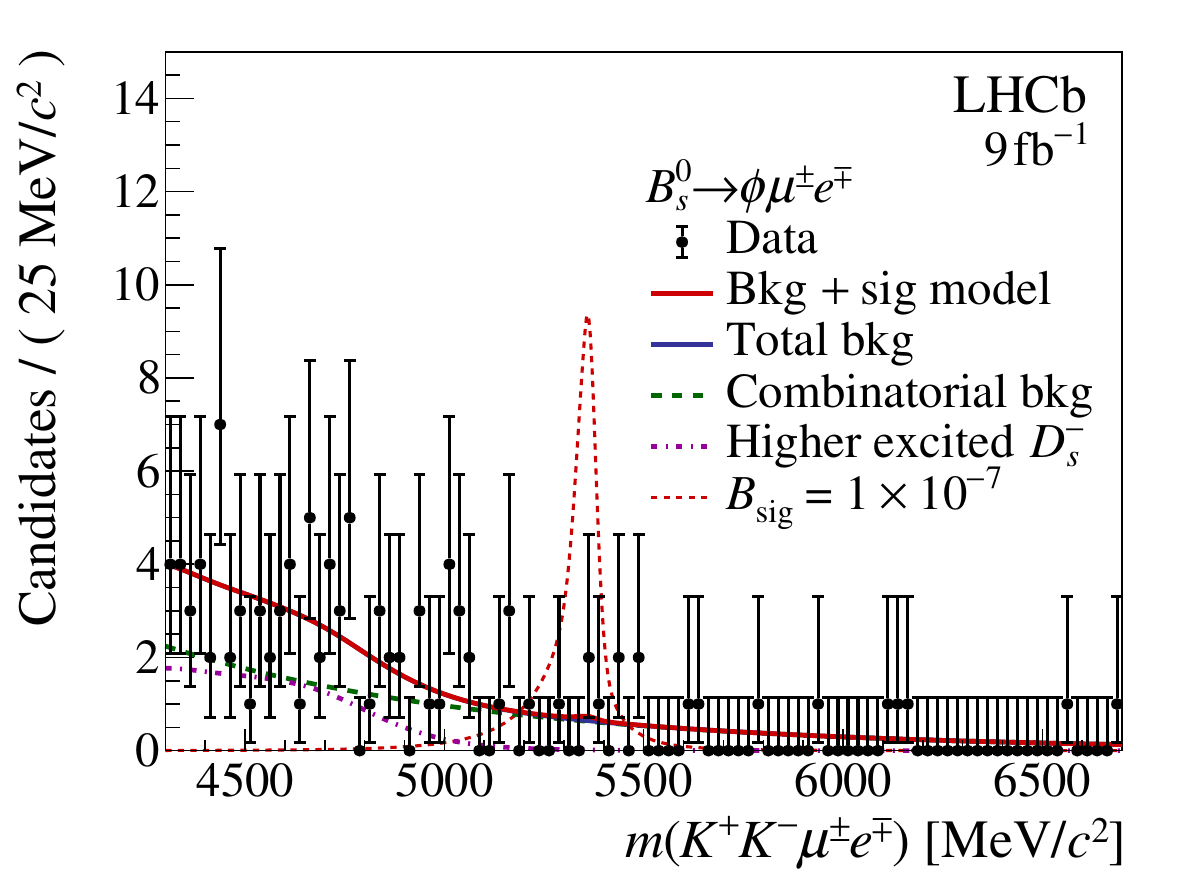}
  \caption{\small
    Mass distributions for (top left)  $\decay{\Bd}{\Kstarz\mu^+ e^-}$, (top right) $\decay{\Bd}{\Kstarz\mu^- e^+}$, (bottom left) $\decay{\Bd}{\Kstarz\mu^\pm e^\mp}$, and (bottom right) $\decay{\Bs}{\phi\mu^\pm e^\mp}$ candidates. The data are overlaid with the fit results.
    For illustration, the signal shape, scaled to a branching fraction of $5\times 10^{-8}$ for the $\decay{\Bd}{\Kstarz\mu^\pm e^\mp}$ decays and $1\times 10^{-7}$ for $\decay{\Bs}{\phi\mu^\pm e^\mp}$, is drawn as red dashed line.\label{fig:massfits}}
\end{figure}

The signal branching fractions are determined in a simultaneous fit of the data taking periods $2011\text{--}2012$, $2015\text{--}2016$, and $2017\text{--}2018$ using Eq.~\ref{eq:bfformula}. 
In the fit, the signal branching fraction and the branching fraction of the 
semileptonic cascade decays involving higher excited $D_{(s)}^-$ resonances, are shared between data taking periods. 
Figure~\ref{fig:massfits} shows the reconstructed $B_{(s)}^0$ mass distributions for $\decay{\Bd}{\Kstarz\mu^\pm e^\mp}$ and $\decay{\Bs}{\phi\mu^\pm e^\mp}$ candidates, combined over data taking periods, 
and overlaid with the fit results for the full and the background-only model. 
For illustration, the signal shape, scaled to a branching fraction of $5\times 10^{-8}$ for the $\decay{\Bd}{\Kstarz\mu^\pm e^\mp}$ decays and $1\times 10^{-7}$ for $\decay{\Bs}{\phi\mu^\pm e^\mp}$, is drawn in addition.
The fits for the same-sign ($\decay{\Bd}{\decay{\Kstarz(}{K^+\pi^-)}\mu^+ e^-}$) and opposite-sign ($\decay{\Bd}{\decay{\Kstarz(}{K^+\pi^-)}\mu^- e^+}$) samples are also given. 
No significant signals are observed and limits on the signal decays are set, as detailed in Sec.~\ref{sec:results}. 

\section{Systematic uncertainties}
\label{sec:systematics}
Systematic effects can modify the limit setting through the normalisation constant $\alpha$ or through modifications of the fit model. 
The systematic effects on the fit model comprise the correction of the signal mass resolution and the exponential slope of the combinatorial background. 
They are included in the fit through Gaussian constraints. 
For the signal mass resolutions, the correction factors for the core Gaussian resolution are allowed to vary within their uncertainties. 
For the exponential slope, the difference between the fit to same-sign lepton data with a reduced BDT cut and with the nominal BDT requirement is used 
as an estimate for the uncertainty on this parameter.
This is added in quadrature to the uncertainty of the parameter from the fit with the reduced BDT requirement, and included as Gaussian constraint in the fit. 

A summary of the systematic uncertainties affecting the normalisation constant $\alpha$ is given in Tab.~\ref{tab:systs}. 
The dominant source of systematic uncertainty originates from the uncertainty on the branching fraction of the normalisation channel. 
For the signal decay $\decay{\Bs}{\phi\mu^\pm e^\mp}$, a systematic uncertainty of similar size arises 
from the significant lifetime difference of the \Bs\ mass eigenstates. 
The effective lifetime of the final state is a priori unknown and can affect the signal efficiency which depends on the \Bs\ decay time. 
The difference between the maximum and minimum lifetimes, given by those of the light and the heavy mass eigenstate, is used to determine a conservative systematic uncertainty.
Further systematic uncertainties arise from the limited size of the simulation samples and the limited precision of the calibration procedures applied to simulation.
These include the weighting of the $B$ production kinematics and event multiplicity,
calibration of the particle identification response and the tracking efficiency, 
as well as the calibration of the trigger efficiencies using data~\cite{LHCb-PUB-2014-039}. 
In addition, systematic uncertainties originating from residual differences between data and simulation 
are conservatively estimated from a comparison of the BDT output distribution for background-subtracted~\cite{Pivk:2004ty} 
\mbox{$\decay{\Bd}{\jpsi(\to\mumu)\Kstarz}$} and \mbox{$\decay{\Bs}{\jpsi(\to\mumu)\phi}$} decays with simulation. 
The relative difference in the normalisation channel BDT selection efficiency in data and simulation is assigned as a systematic uncertainty.

\begin{table}
\caption{Sources of relative systematic uncertainties $[\%]$ on the normalisation constant $\alpha$ defined in Eq.~\ref{eq:bfformula}. Where the uncertainty depends on the year of data taking, a range is provided.\label{tab:systs}}
  \centering
  \begin{tabular}{lrr}\hline
	Systematic source 					& $\decay{\Bd}{\Kstarz\mu^\pm e^\mp}$ & $\decay{\Bs}{\phi\mu^\pm e^\mp}$\\\hline
	Normalisation $\mathcal{B}$			& $4.0$ 						& $4.8$						\\\hline
	$\phi\mu^\pm e^\mp$ decay time distribution							& --								& $3.8\text{--}4.5$			\\ 
	Limited simulation sample size						& $0.7\text{--}1.5$			& $0.6\text{--}1.4$			\\
	Production kinematics/multiplicity						& $1.2\text{--}2.7$			& $1.6\text{--}3.8$			\\
	Particle identification							& $0.3\text{--}0.8$			& $0.3\text{--}0.6$			\\
	Muon identification 							& $1$						& $1$						\\
	Tracking efficiency						& $1$						& $1$						\\
	\texttt{L0} trigger calibration			& $1$						& $1$						\\
	\texttt{HLT} trigger efficiency			& $1$						& $1$						\\
	Residual MC differences								& $1.7\text{--}4.3$			& $1.1\text{--}3.7$			\\\hline
	Sum						& $5.2\text{--}6.7$	& $6.9\text{--}8.5$	\\
	\hline\end{tabular}  
\end{table}

\section{Results}
\label{sec:results}
No significant excess of $\decay{\Bd}{\Kstarz\mu^\pm e^\mp}$ or $\decay{\Bs}{\phi\mu^\pm e^\mp}$ decays is observed and limits are set using the \cls\ method~\cite{CLs}. 
A one-sided test statistic is used~\cite{Cowan:2010js}, as implemented in the \textsc{GammaCombo} framework~\cite{GammaCombo,LHCb-PAPER-2016-032}. 
The test statistic is evaluated using pseudoexperiments, 
which are generated using the best fit values for the nuisance parameters, and where the central values of the Gaussian constraints are varied according to their uncertainties. 

\begin{figure}
  \centering
  \includegraphics[width=0.48\textwidth]{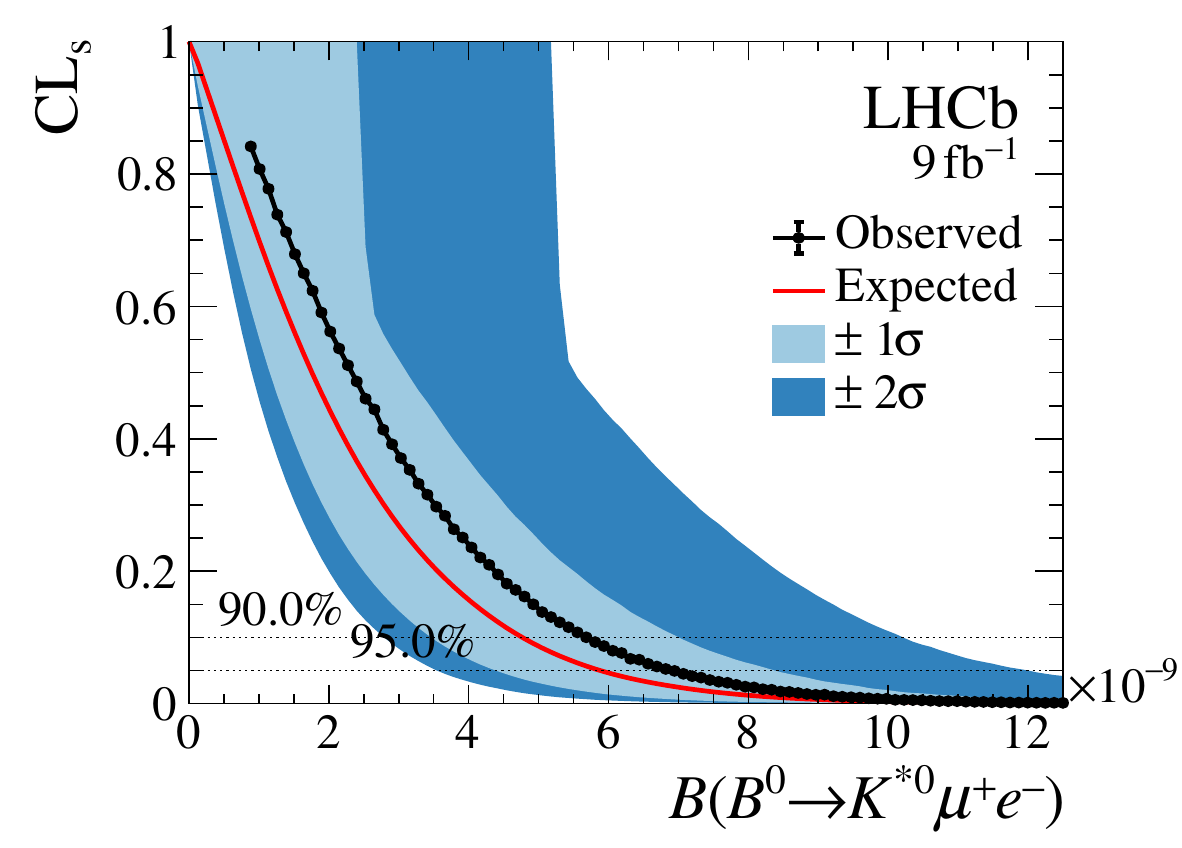}
  \includegraphics[width=0.48\textwidth]{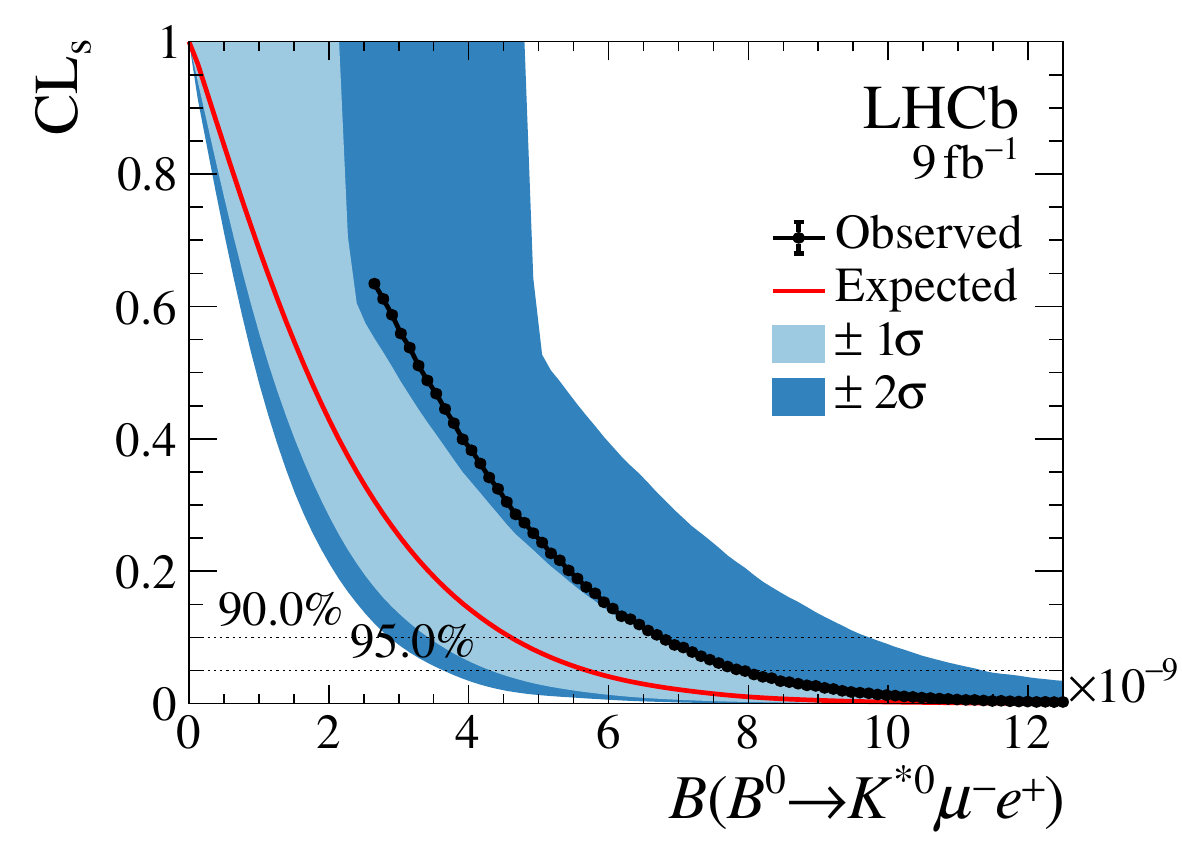}\\
  \includegraphics[width=0.48\textwidth]{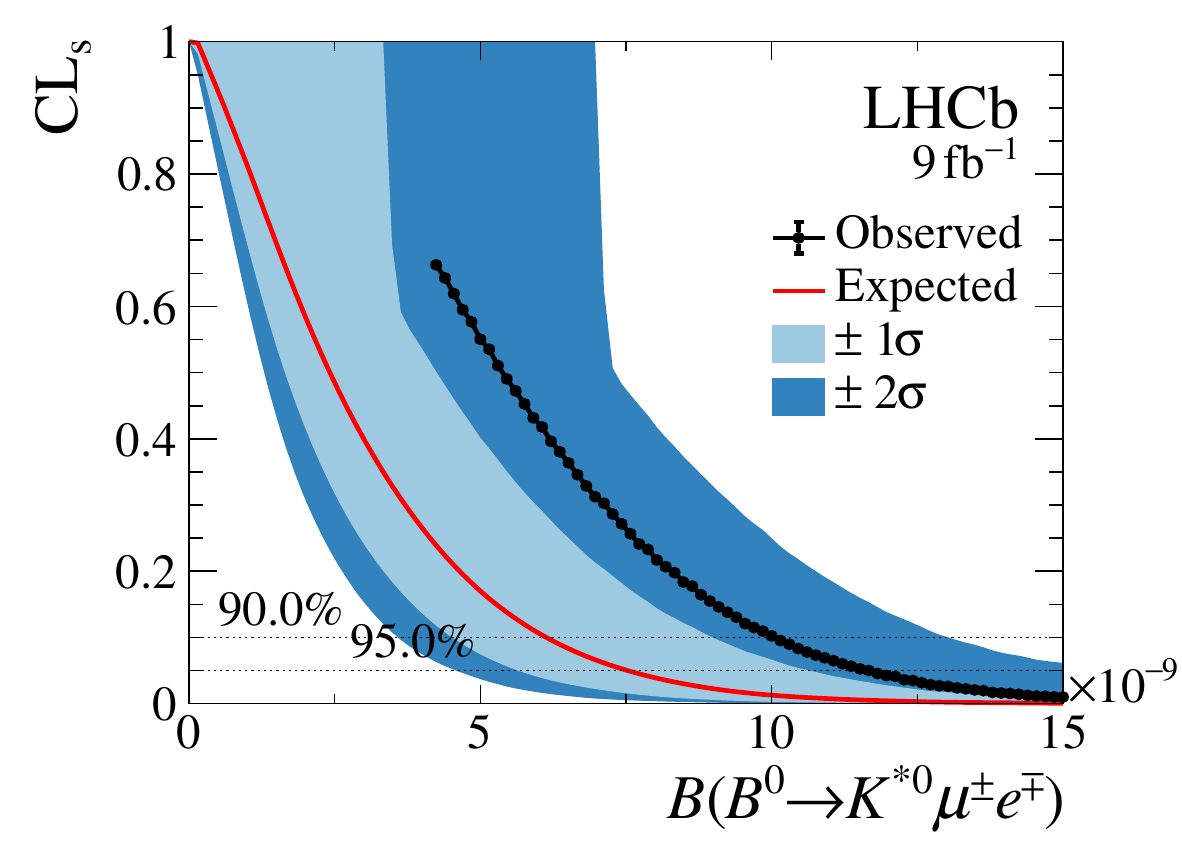}
  \includegraphics[width=0.48\textwidth]{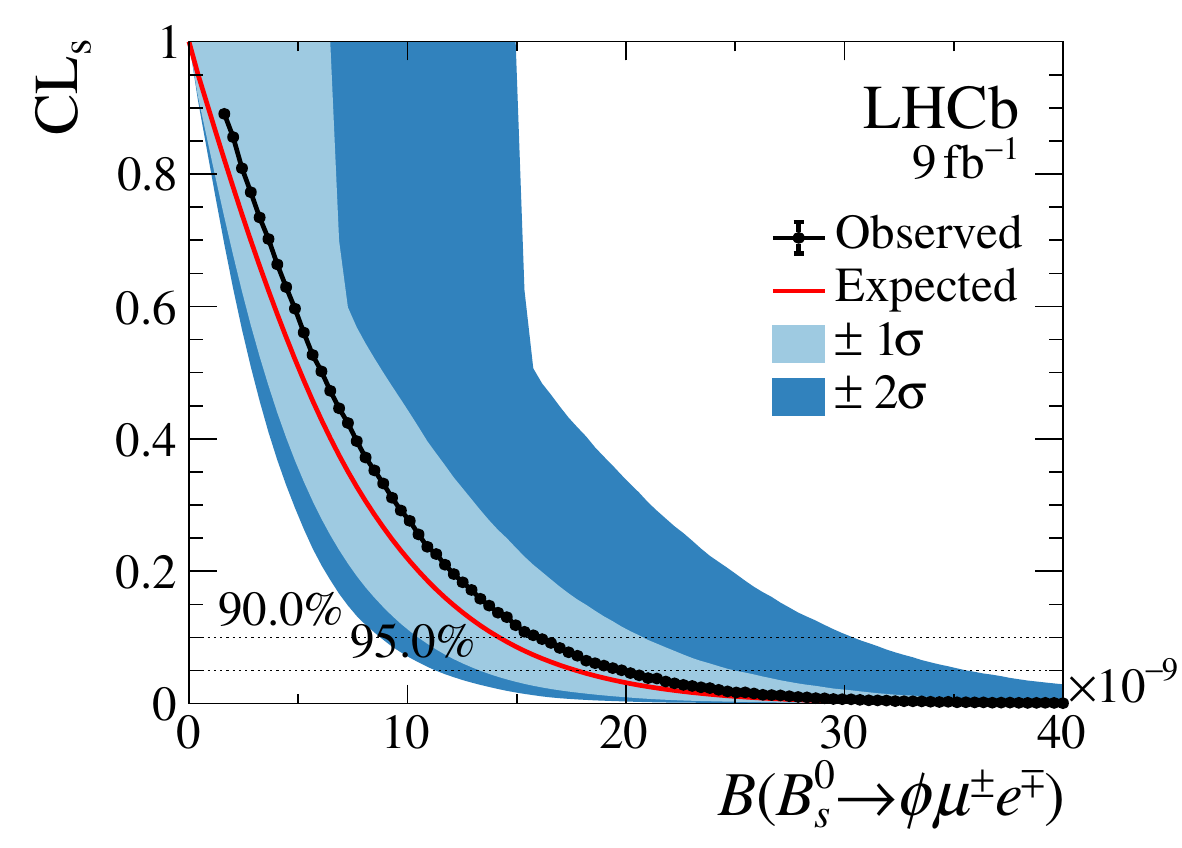}
  \caption{\small
    Observed and expected (background-only hypothesis) limits for (top left) $\decay{\Bd}{\Kstarz\mu^+ e^-}$, (top right) $\decay{\Bd}{\Kstarz\mu^- e^+}$,
    (bottom left) $\decay{\Bd}{\Kstarz\mu^\pm e^\mp}$, and (bottom right) $\decay{\Bs}{\phi\mu^\pm e^\mp}$.\label{fig:limits}}
\end{figure}

The resulting \cls\ scans are shown in Fig.~\ref{fig:limits}, and upper limits at $90\%$ and $95\%$ CL are reported in Tab.~\ref{tab:limits};
the limits given for $\BF(\decay{\Bd}{\Kstarz\mu^\pm e^\mp})$ are determined using the combined $\decay{\Bd}{\Kstarz\mu^+ e^-}$ and $\decay{\Bd}{\Kstarz\mu^- e^+}$ data samples.
\begin{table}
  \caption{\small
    Expected (background-only hypothesis) and observed limits $[10^{-9}]$ at $90\%$ ($95\%$) CL.\label{tab:limits}
  }
  \centering
\begin{tabular}{lcc}
  \hline
  Mode & Expected & Observed\\ \hline
$\decay{\Bd}{\Kstarz\mu^+ e^-}$        &  $4.8$ ($5.9$)    &  $5.7$ ($6.9$)    \\
$\decay{\Bd}{\Kstarz\mu^- e^+}$        &  $4.6$ ($5.7$)    &  $6.8$ ($7.9$)    \\
$\decay{\Bd}{\Kstarz\mu^\pm e^\mp}$    &  $6.1$ ($7.5$)    &  $10.1$ ($11.7$)  \\
$\decay{\Bs}{\phi\mu^\pm e^\mp}$       &  $14.2$ ($17.7$)  &  $16.0$ ($19.8$)  \\
\hline\end{tabular}
\end{table}

The default limits assume a uniform phase space model for the signal decays. 
However, NP models can result in very different 
decay kinematics and differential decay rates~\cite{Becirevic:2016zri,Altmannshofer:2008dz,Bobeth:2008ij}.
This is illustrated in Fig.~\ref{fig:npmodels} of App.~\ref{app:npmodels}
for two distinct NP scenarios:
a scalar model with $C_S^{\mu e}\neq 0$,
and a left-handed model with $C_9^{\mu e}=-C_{10}^{\mu e}\neq 0$~\cite{Straub:2018kue,Becirevic:2016zri}. 
Here, $C_i^{\mu e}$ denotes the lepton-flavour violating Wilson coefficients. 
As the reconstruction and selection efficiency shown in Fig.~\ref{fig:efficiencies} in App.~\ref{app:efficiencies} is not flat in
the decay kinematics,
the total signal efficiency can differ for NP models, which needs to be accounted for in the computation of limits.
For the scalar- and left-handed NP scenarios shown in App.~\ref{app:npmodels}, 
the resulting limits are given in Tab.~\ref{tab:nplimits}. 
\begin{table}
  \caption{\small
    Exclusion limits $[10^{-9}]$ on the $B_{(s)}^0$ branching fractions for a scalar ($C_s^{\mu e}\neq 0$) and left-handed ($C_9^{\mu e}=-C_{10}^{\mu e}\neq 0$) NP model at $90\%$ ($95\%$) CL.\label{tab:nplimits}}
  \centering
\begin{tabular}{lrr}\hline
	 Mode & \multicolumn{1}{c}{Left-handed} & \multicolumn{1}{c}{Scalar} \\ \hline
	$\decay{\Bd}{\Kstarz\mup\en}$ 	&  $6.7$ ($8.3$) &  $8.4$ ($10.2$) \\
	$\decay{\Bd}{\Kstarz\mun\ep}$ 	&  $8.0$ ($9.5$) &  $9.9$ ($11.5$) \\
	$\decay{\Bd}{\Kstarz\mupm\emp}$ &  $12.0$ ($13.9$) &  $14.7$ ($17.0$) \\
	$\decay{\Bs}{\phiz\mupm\emp}$	&  $16.5$ ($20.5$) &  $18.8$ ($23.1$) \\
\hline\end{tabular}
\end{table}

\section{Conclusions}
\label{sec:conclusions}
A search for the lepton-flavour violating decays $\decay{\Bd}{\Kstarz\mu^+ e^-}$, $\decay{\Bd}{\Kstarz\mu^- e^+}$, $\decay{\Bd}{\Kstarz\mu^\pm e^\mp}$, and $\decay{\Bs}{\phi\mu^\pm e^\mp}$ is presented. 
The search uses $pp$ collision data collected
at centre-of-mass energies of $7\tev$ ($2011$), $8\tev$ ($2012$), and $13\tev$ ($2015\text{--}2018$), 
cor\-responding to a total integrated luminosity of $9\invfb$. 
No significant excesses are observed and limits on the branching fractions are obtained 
for a uniform phase space signal decay model, resulting in
\begin{align*}
    {\cal B}(\decay{\Bd}{\Kstarz\mu^+ e^-}) &< \phantom{1}5.7\times 10^{-9}~(6.9\times 10^{-9}),\\
    {\cal B}(\decay{\Bd}{\Kstarz\mu^- e^+}) &< \phantom{1}6.8\times 10^{-9}~(7.9\times 10^{-9}),\\
    {\cal B}(\decay{\Bd}{\Kstarz\mu^\pm e^\mp}) &< 10.1\times 10^{-9}~(11.7\times 10^{-9}), \\
    {\cal B}(\decay{\Bs}{\phi\mu^\pm e^\mp}) &< 16.0\times 10^{-9}~(19.8\times 10^{-9})
\end{align*}
at $90\%$ ($95\%$) CL. 
In addition, limits on a scalar and left-handed NP scenarios are reported. 
The results constitute the most stringent limits on a semileptonic LFV $b$-hadron decays to date. 
The limits on the decay $\decay{\Bd}{\Kstarz\mu^\pm e^\mp}$ are improved by more than one order of magnitude in comparison to previous searches. 
The reported limits on the decay $\decay{\Bs}{\phi\mu^\pm e^\mp}$ constitute the world's first constraint of a semileptonic LFV \Bs\ decay. 

\nocite{LHCb-PAPER-2013-019}

\section*{Acknowledgements}
%
%
\noindent We express our gratitude to our colleagues in the CERN
accelerator departments for the excellent performance of the LHC. We
thank the technical and administrative staff at the LHCb
institutes.
We acknowledge support from CERN and from the national agencies:
CAPES, CNPq, FAPERJ and FINEP (Brazil); 
MOST and NSFC (China); 
CNRS/IN2P3 (France); 
BMBF, DFG and MPG (Germany); 
INFN (Italy); 
NWO (Netherlands); 
MNiSW and NCN (Poland); 
MEN/IFA (Romania); 
MICINN (Spain); 
SNSF and SER (Switzerland); 
NASU (Ukraine); 
STFC (United Kingdom); 
DOE NP and NSF (USA).
We acknowledge the computing resources that are provided by CERN, IN2P3
(France), KIT and DESY (Germany), INFN (Italy), SURF (Netherlands),
PIC (Spain), GridPP (United Kingdom), 
CSCS (Switzerland), IFIN-HH (Romania), CBPF (Brazil),
Polish WLCG  (Poland) and NERSC (USA).
We are indebted to the communities behind the multiple open-source
software packages on which we depend.
Individual groups or members have received support from
ARC and ARDC (Australia);
Minciencias (Colombia);
AvH Foundation (Germany);
EPLANET, Marie Sk\l{}odowska-Curie Actions and ERC (European Union);
A*MIDEX, ANR, IPhU and Labex P2IO, and R\'{e}gion Auvergne-Rh\^{o}ne-Alpes (France);
Key Research Program of Frontier Sciences of CAS, CAS PIFI, CAS CCEPP, 
Fundamental Research Funds for the Central Universities, 
and Sci. \& Tech. Program of Guangzhou (China);
GVA, XuntaGal, GENCAT and Prog. Atracci\'on Talento, CM (Spain);
SRC (Sweden);
the Leverhulme Trust, the Royal Society
 and UKRI (United Kingdom).

\addcontentsline{toc}{section}{References}
\bibliographystyle{LHCb}
\bibliography{main,standard,LHCb-PAPER,LHCb-CONF,LHCb-DP,LHCb-TDR}
 
\clearpage

\section*{Appendices}

\appendix

\section{New Physics models}
\label{app:npmodels}

\begin{figure}[htbp]
  \centering
  \includegraphics[width=0.36\textwidth]{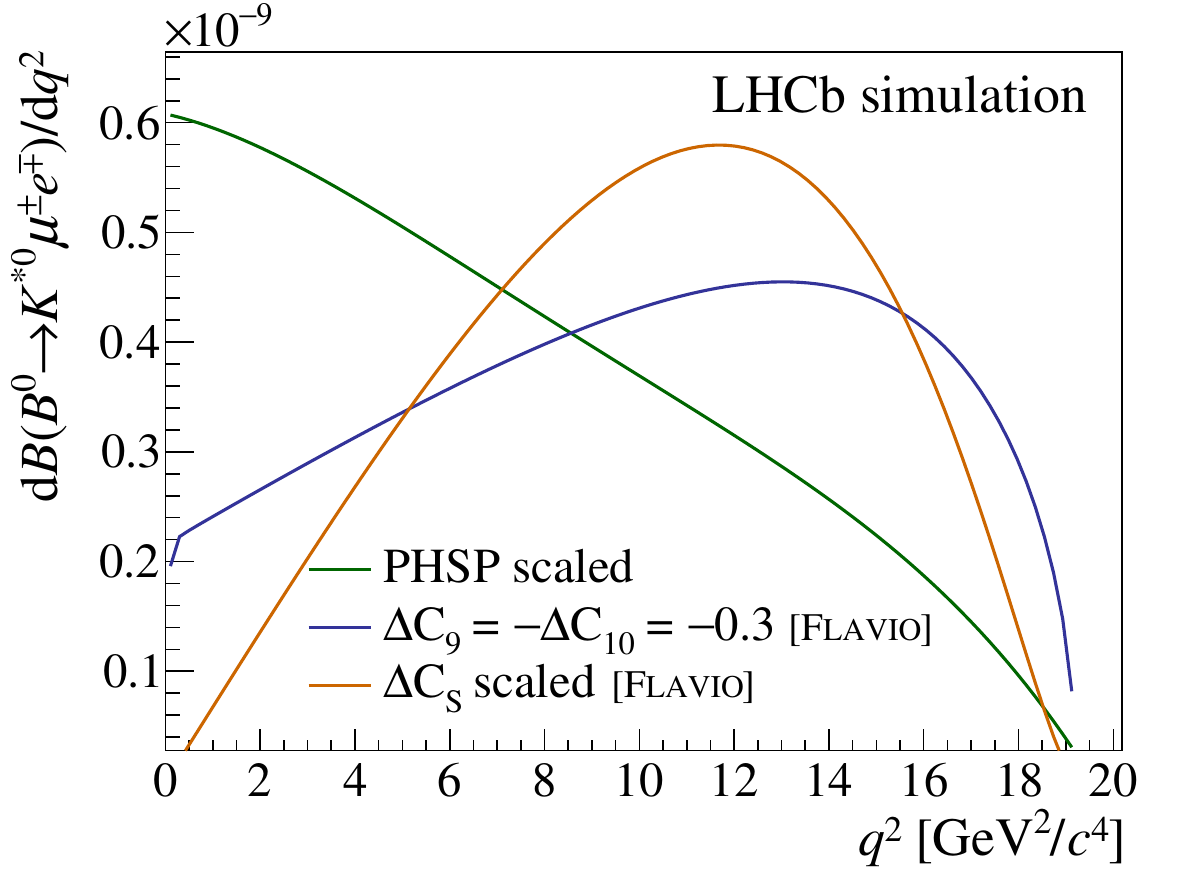}
  \includegraphics[width=0.36\textwidth]{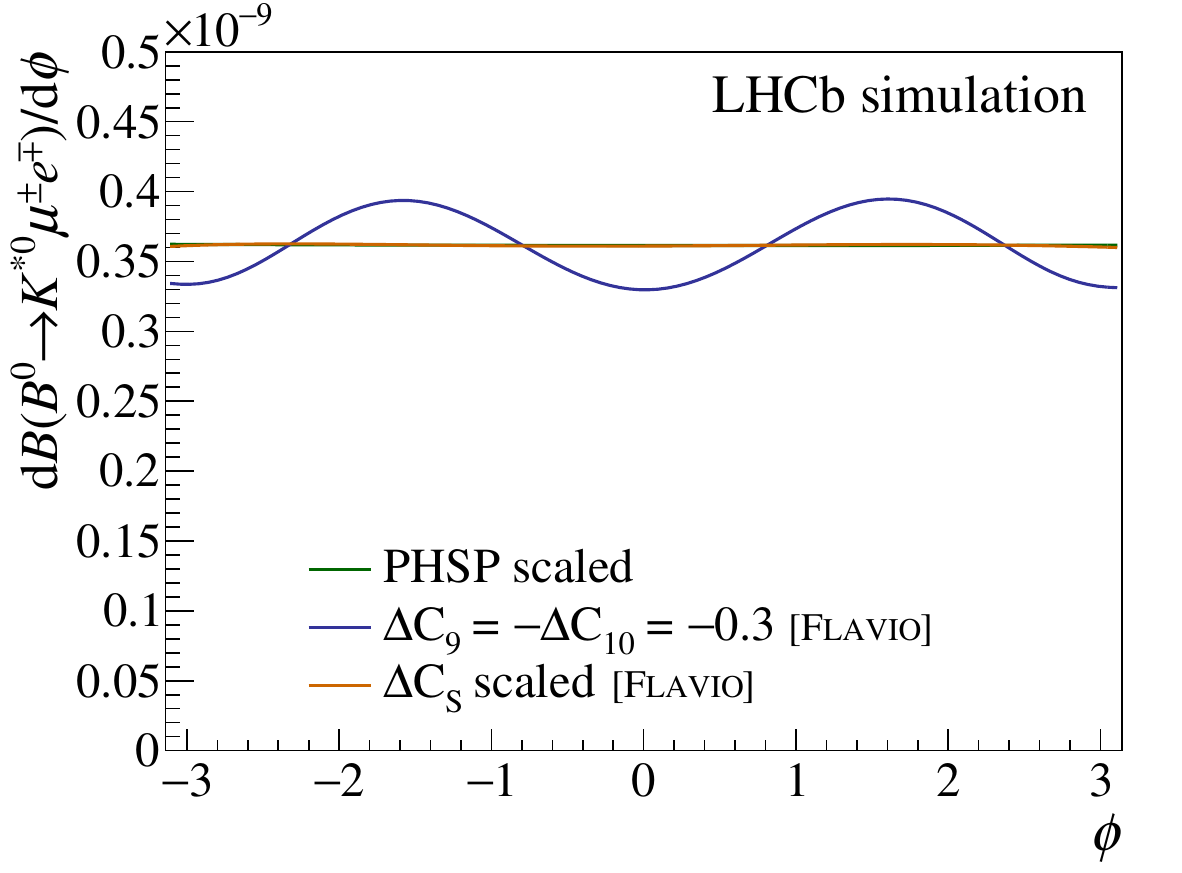}\\
  \includegraphics[width=0.36\textwidth]{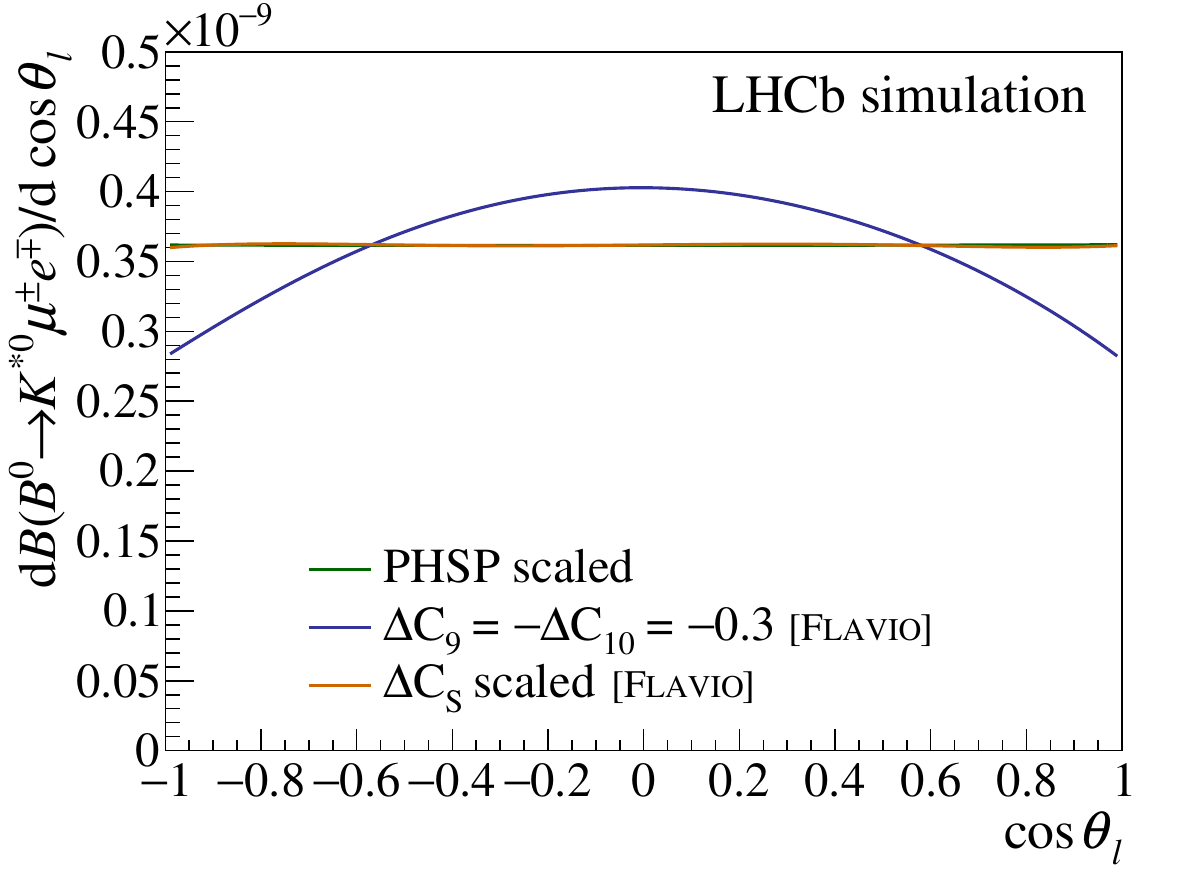}
  \includegraphics[width=0.36\textwidth]{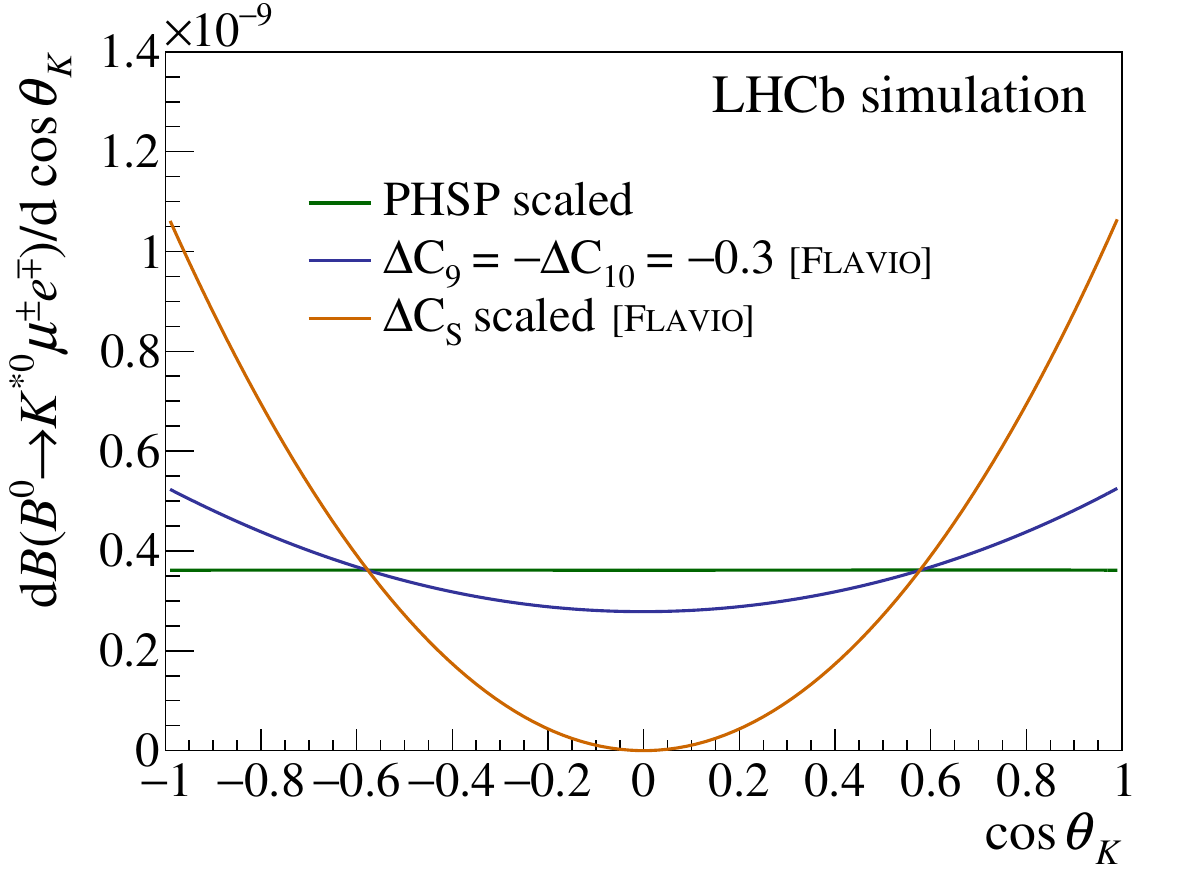}\\
  \includegraphics[width=0.36\textwidth]{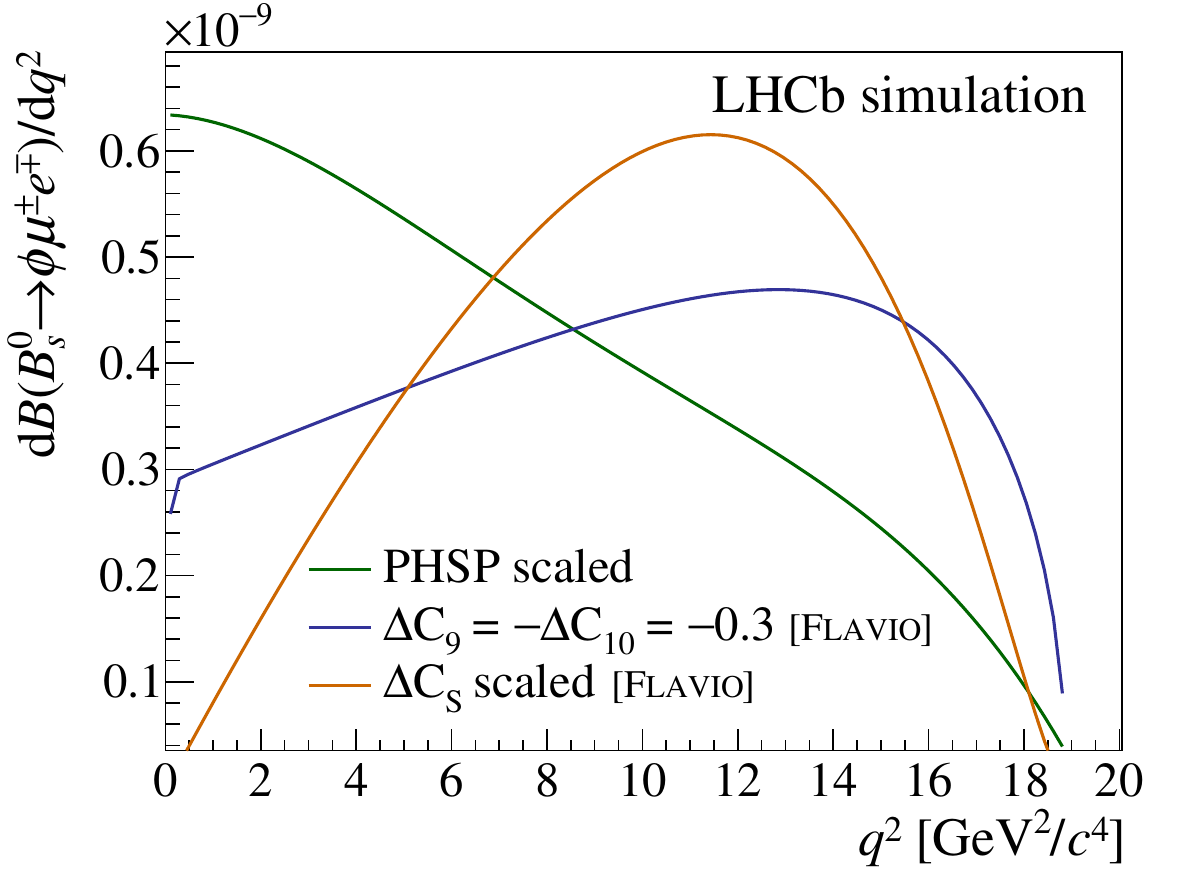}
  \includegraphics[width=0.36\textwidth]{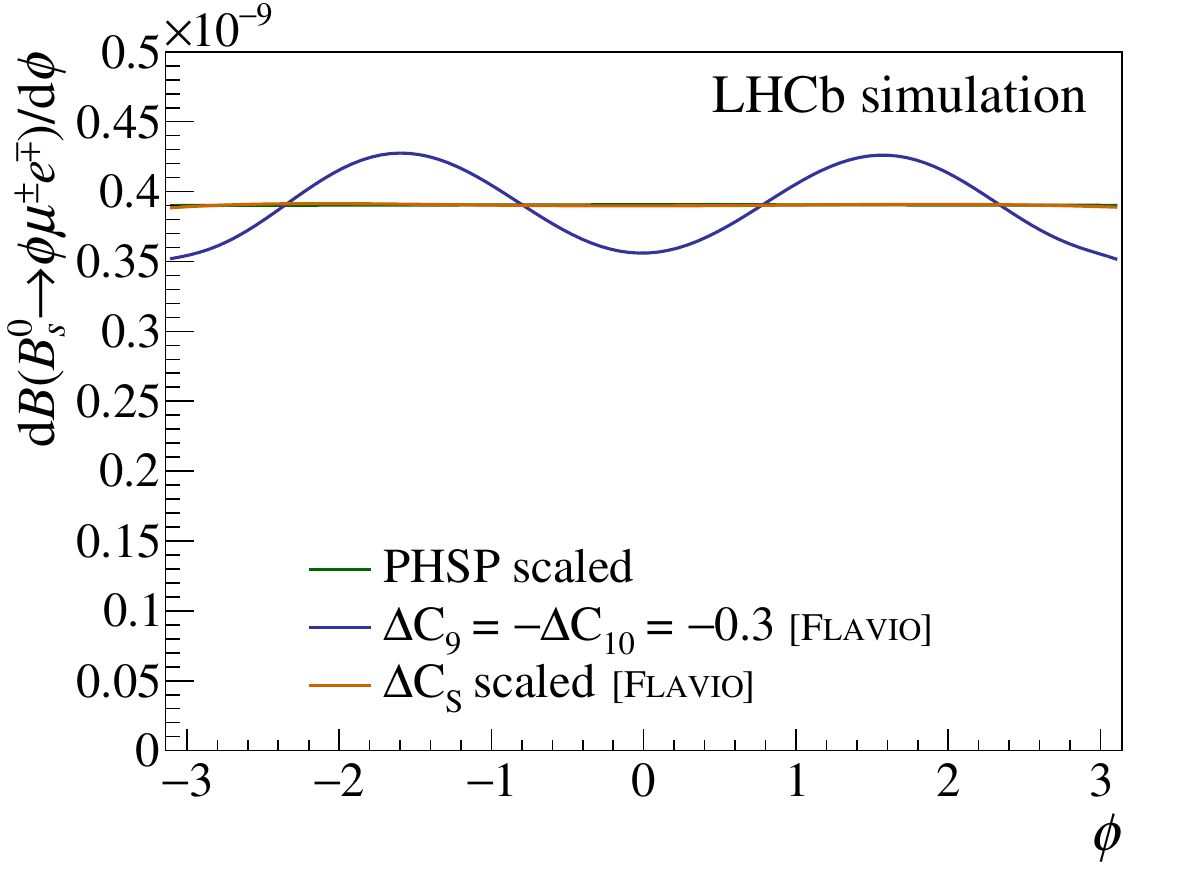}\\
  \includegraphics[width=0.36\textwidth]{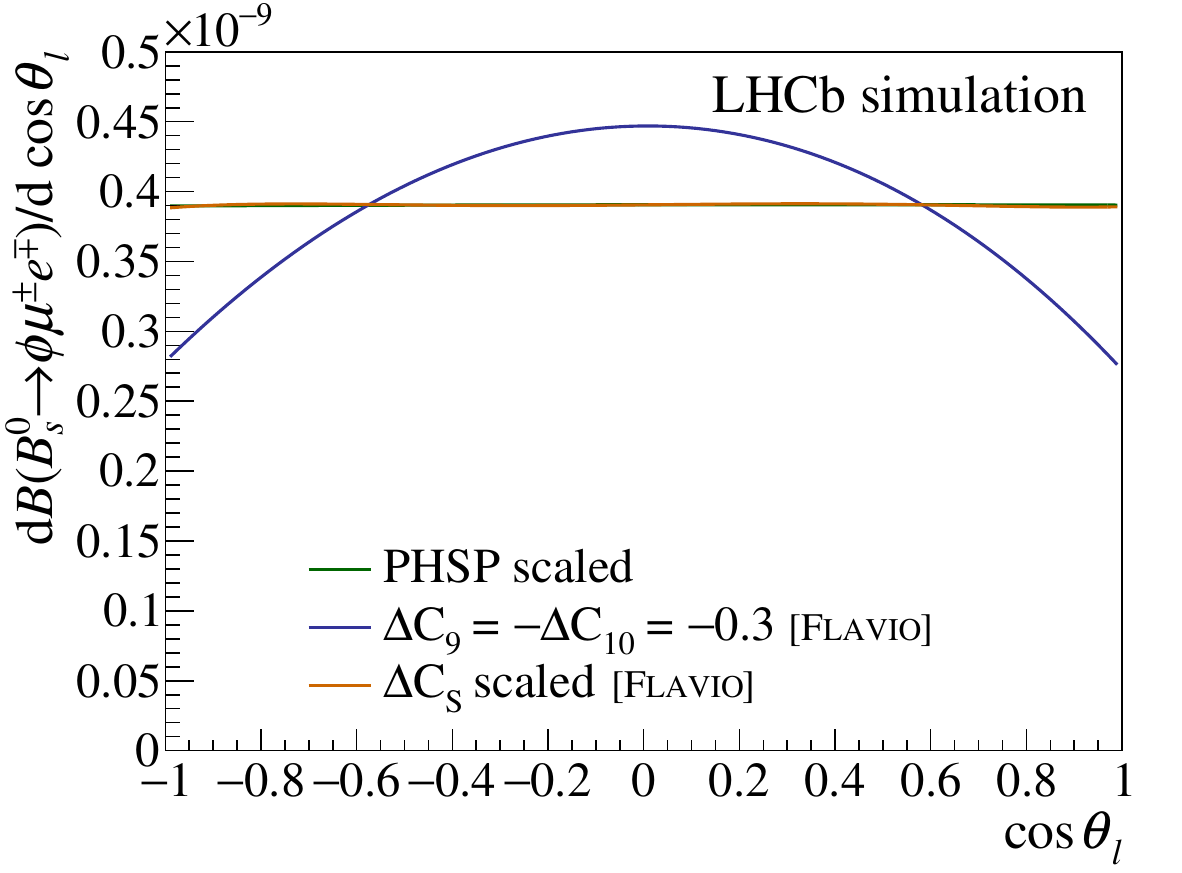}
  \includegraphics[width=0.36\textwidth]{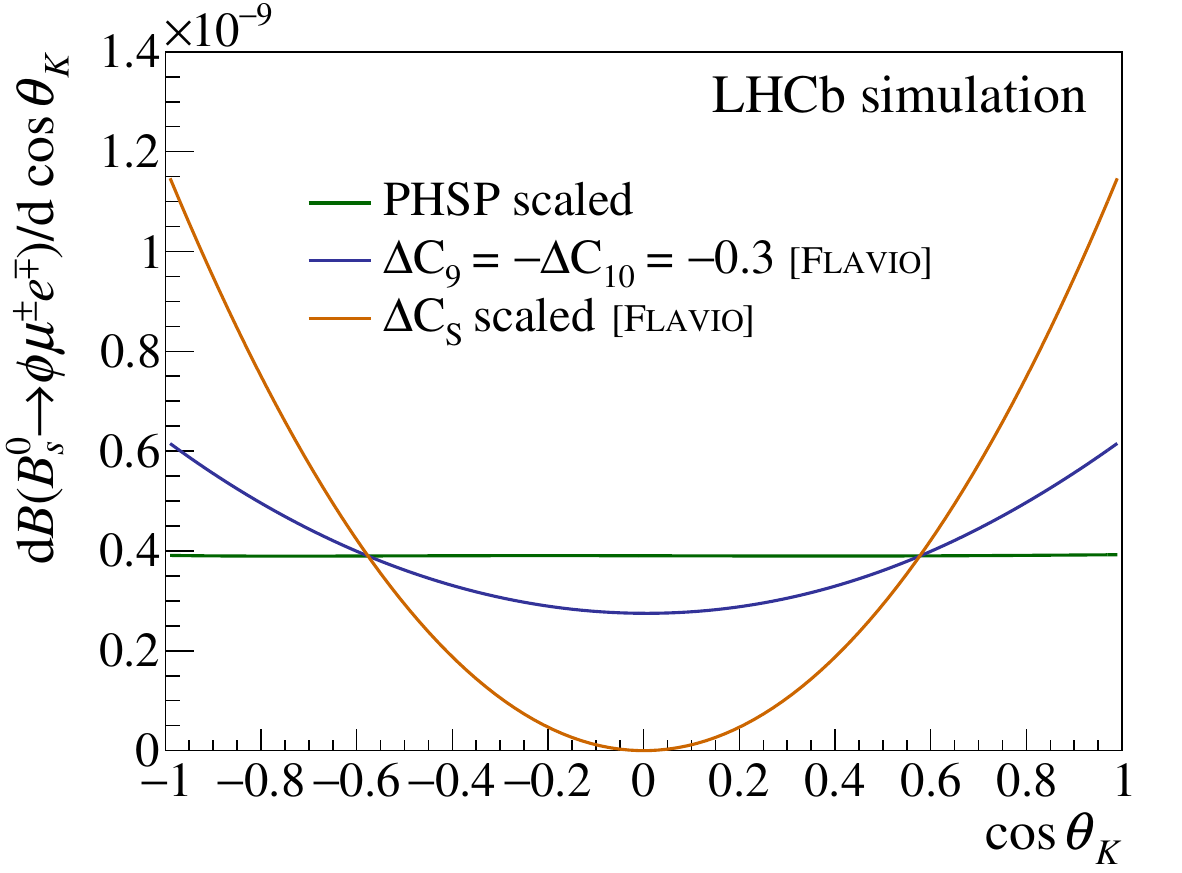}\\
  \caption{\small
Differential decay rate as a function of the four-momentum transfer \qsq\ and the three decay angles in a left-handed ($C_9^{\mu e}=-C_{10}^{\mu e}\neq 0$) NP model for (top) the signal decay $\decay{\Bd}{\Kstarz\mu^\pm e^\mp}$ and (bottom) $\decay{\Bs}{\phi\mu^\pm e^\mp}$.
The left-handed NP scenario is compared with the nominal phase space and a scalar ($C_S^{\mu e}\neq 0$) model, normalised to the same area.
\label{fig:npmodels}}
\end{figure}

New Physics scenarios can result in very different distributions in the three decay angles $\ctl$, $\ctk$ and $\phi$, defined as in Ref.~\cite{LHCb-PAPER-2013-019}, and the four-momentum transfer $\qsq=m^2(\mu^\pm e^\mp)$~\cite{Becirevic:2016zri,Altmannshofer:2008dz,Bobeth:2008ij}.
This is illustrated in Fig.~\ref{fig:npmodels}.

\clearpage

\section{Efficiency projections}
\label{app:efficiencies}

\begin{figure}[htbp]
  \centering
  \includegraphics[width=0.4\textwidth]{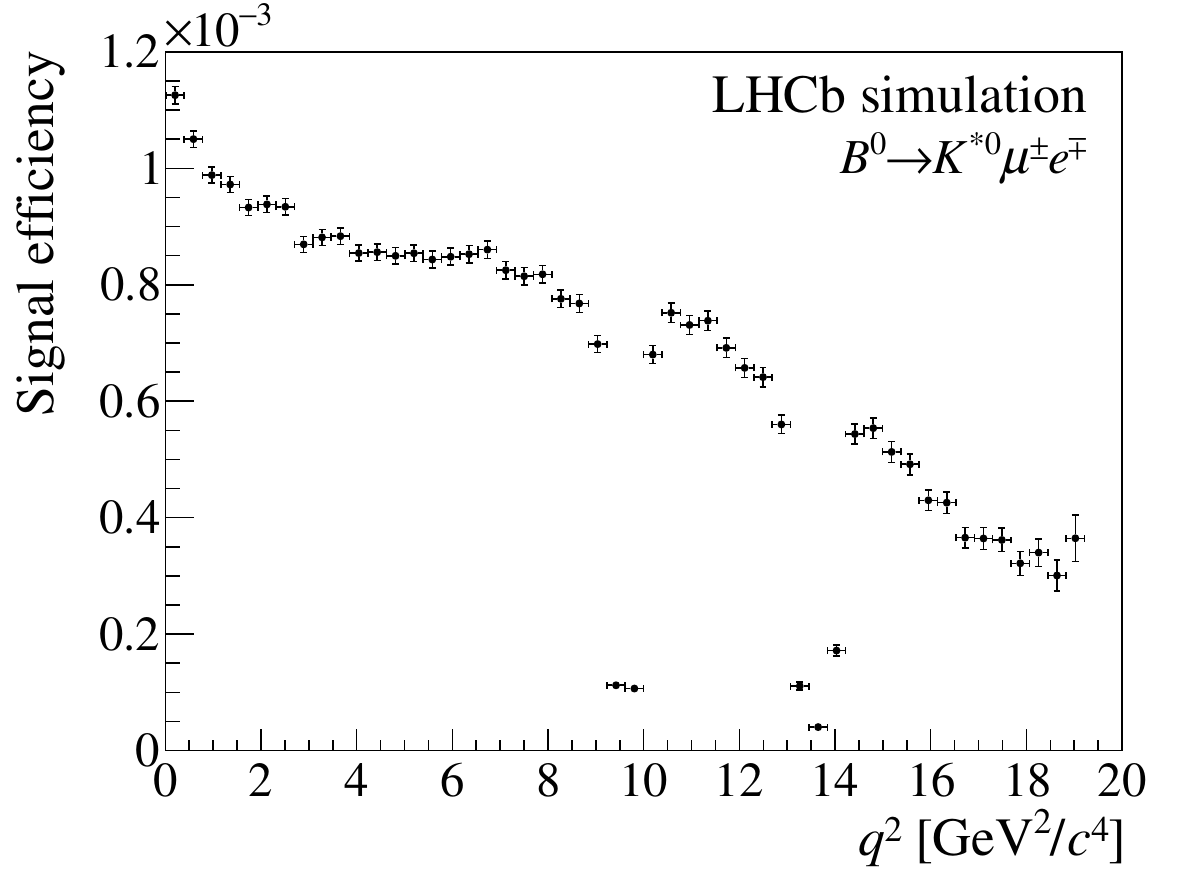}
  \includegraphics[width=0.4\textwidth]{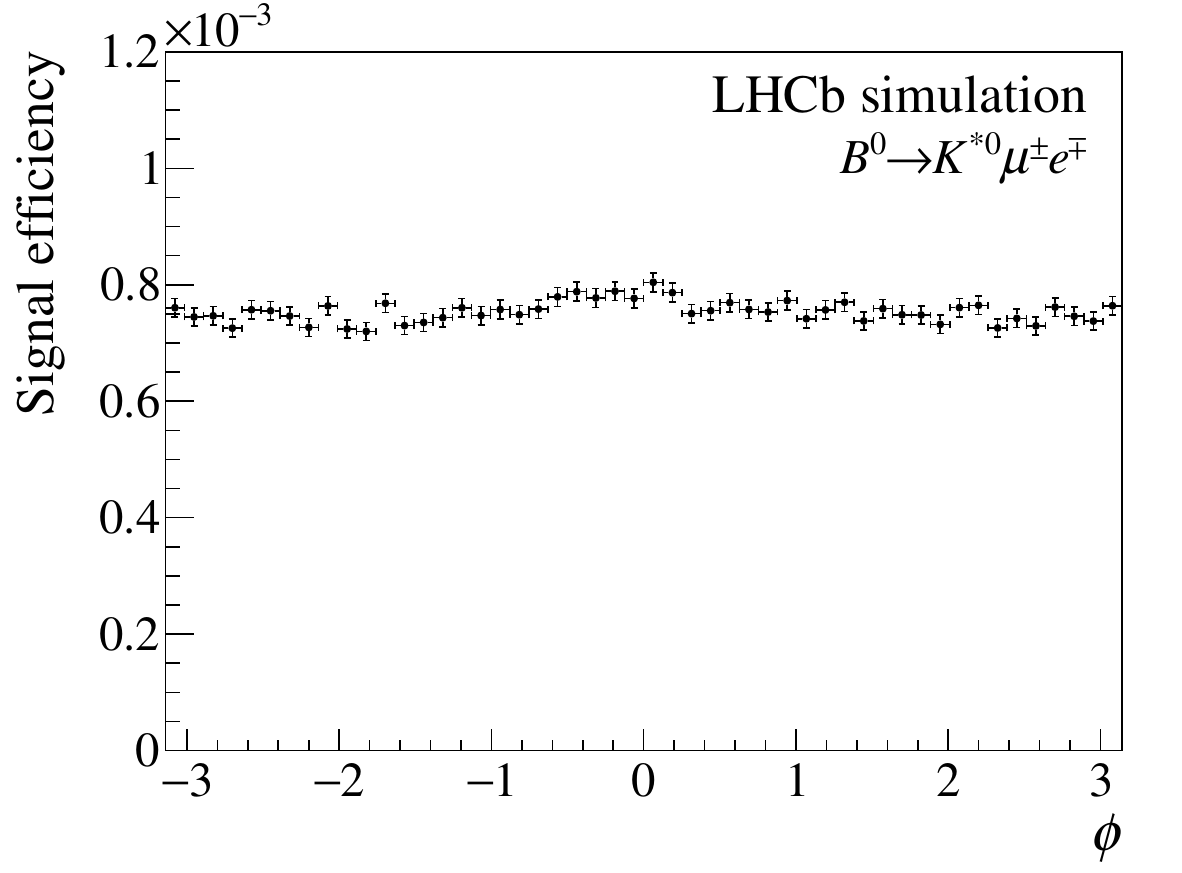}\\
  \includegraphics[width=0.4\textwidth]{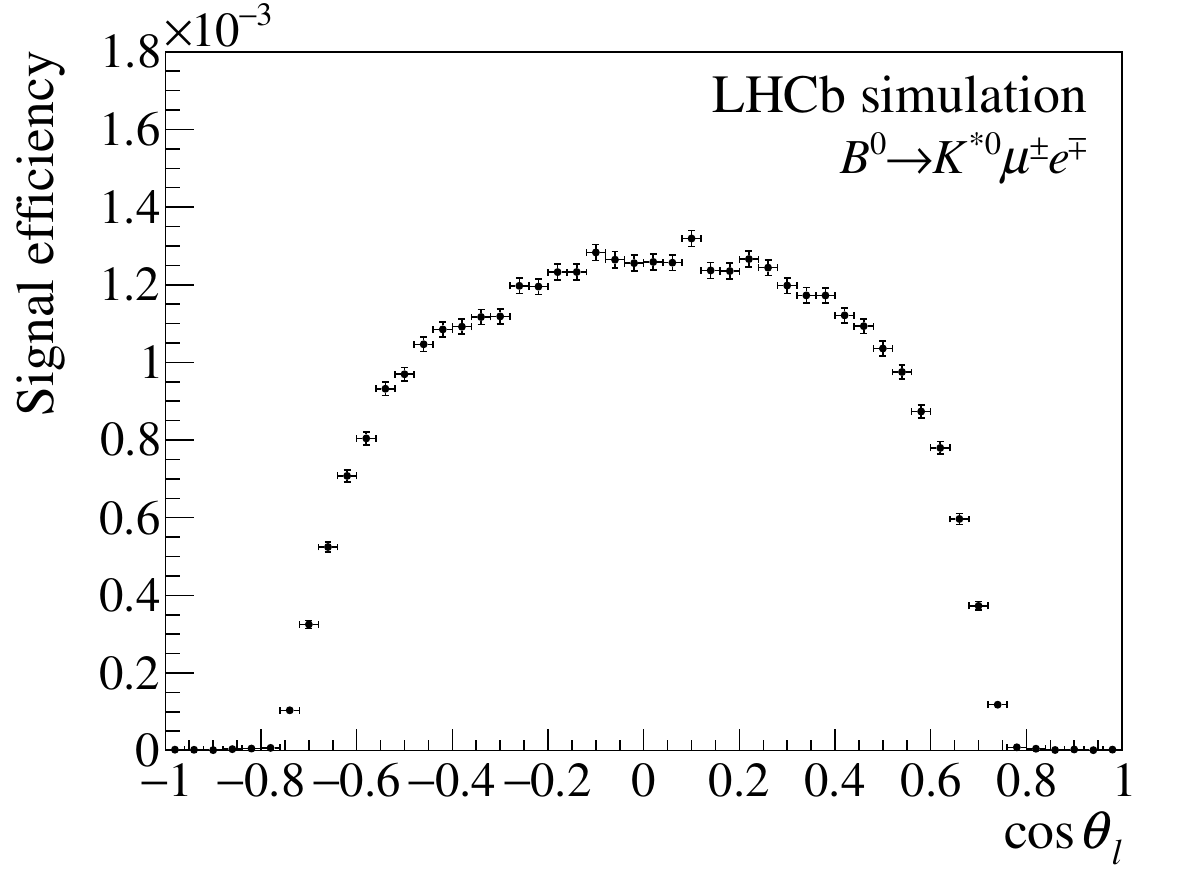}
  \includegraphics[width=0.4\textwidth]{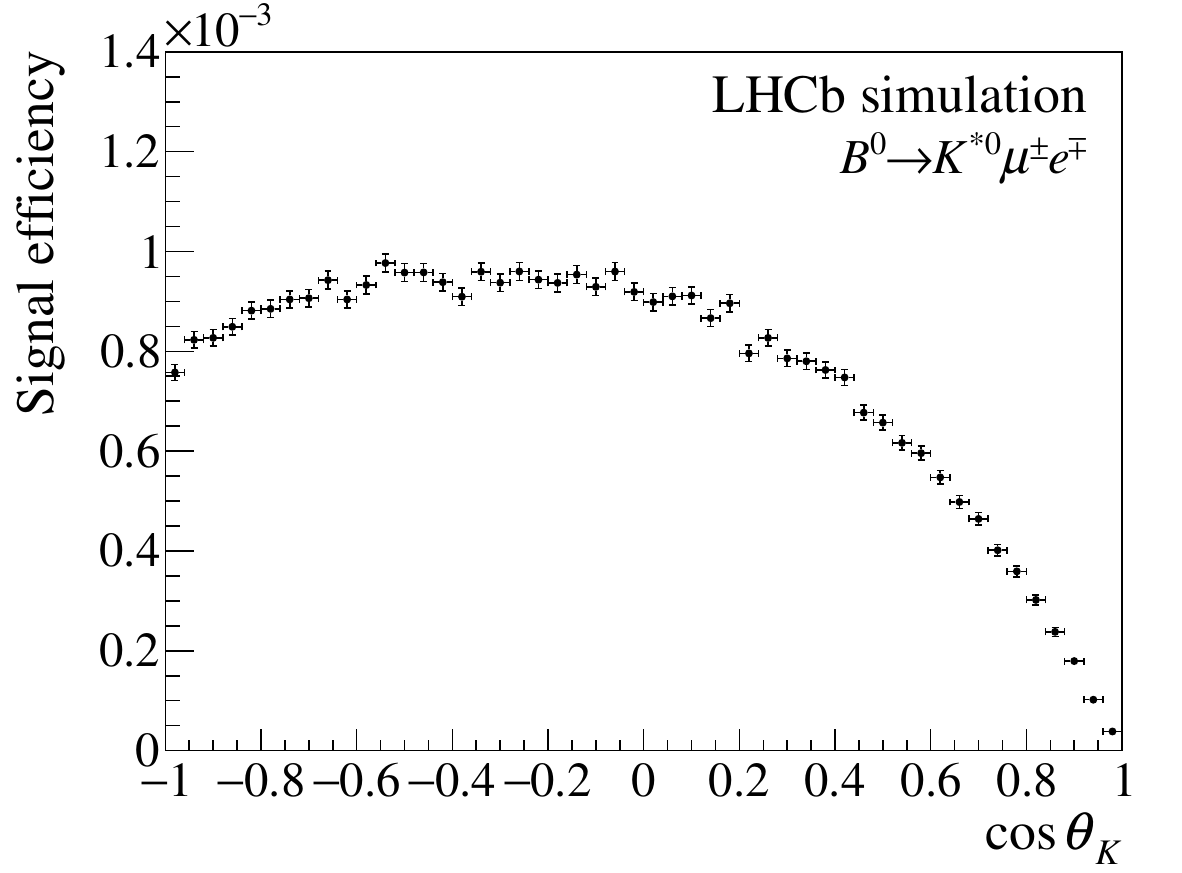}\\
  \includegraphics[width=0.4\textwidth]{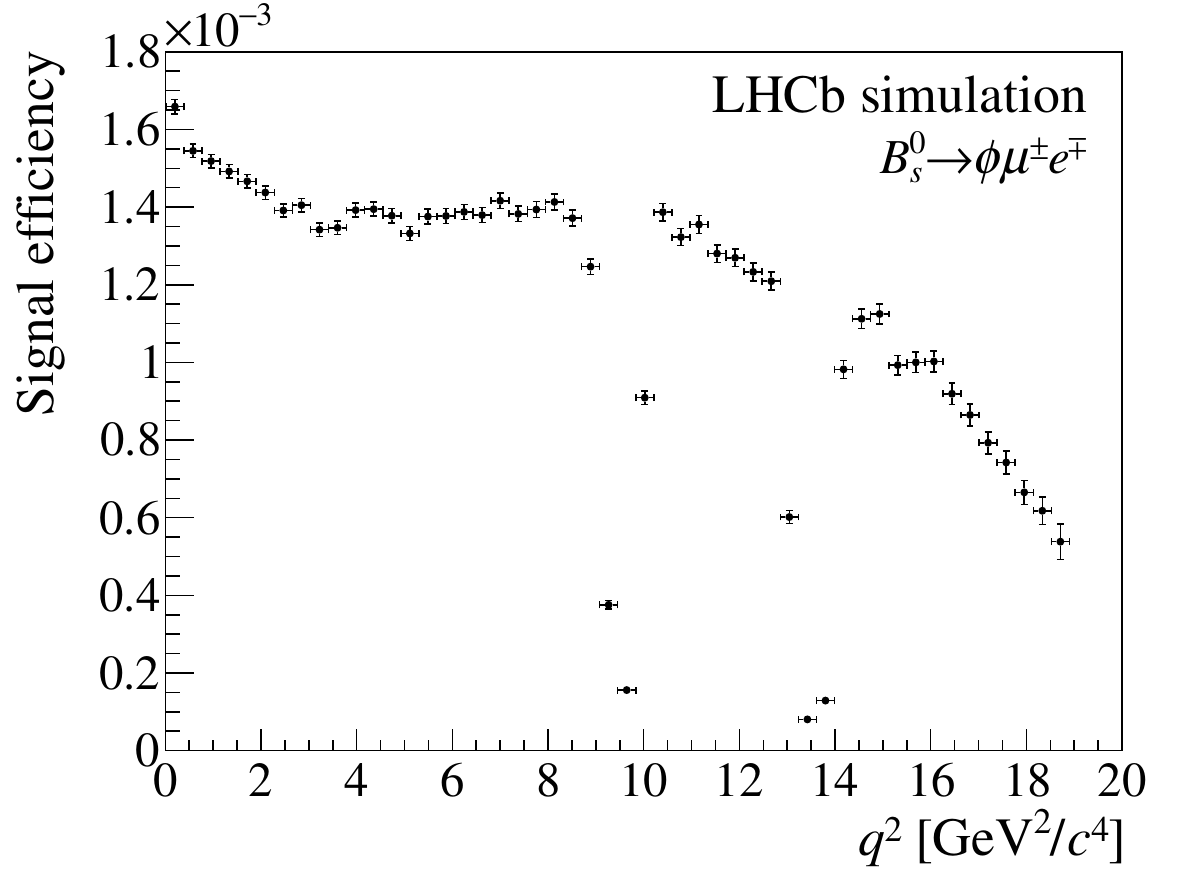}
  \includegraphics[width=0.4\textwidth]{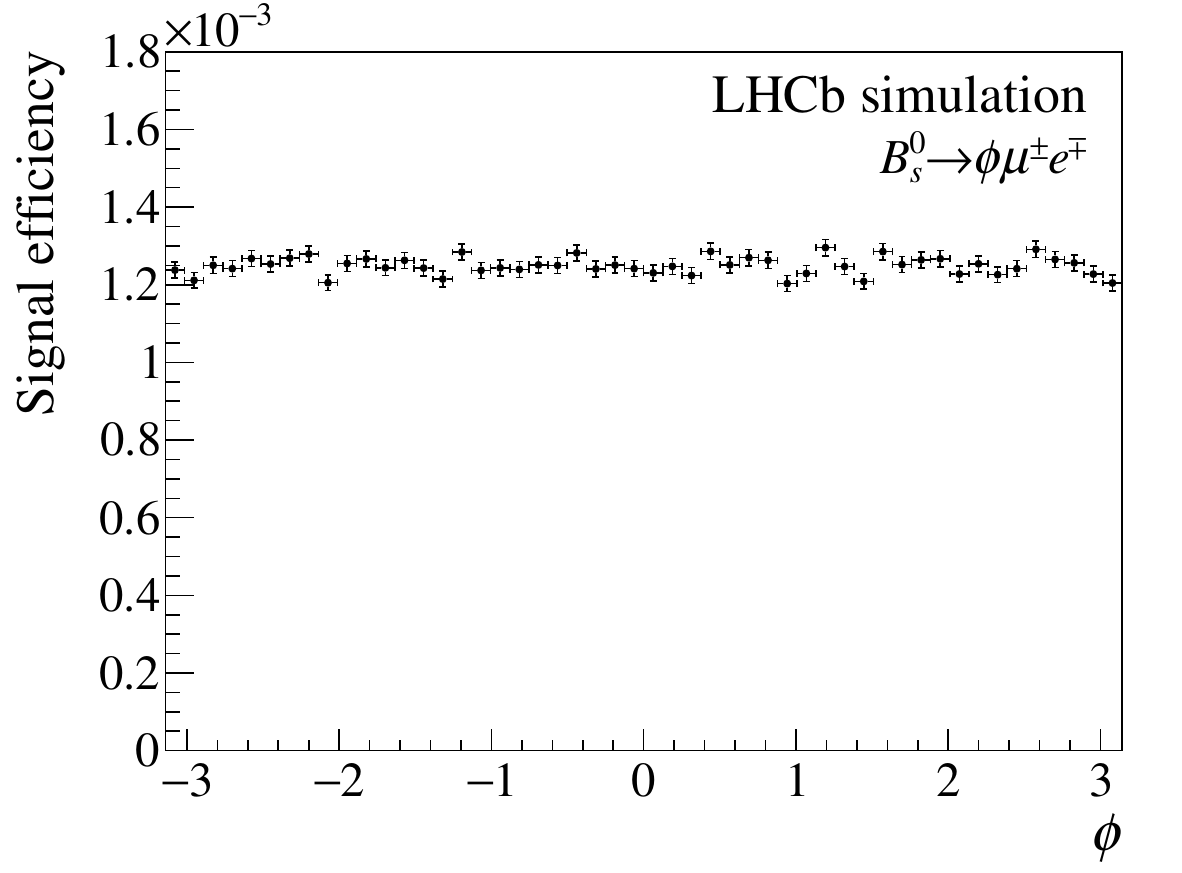}\\
  \includegraphics[width=0.4\textwidth]{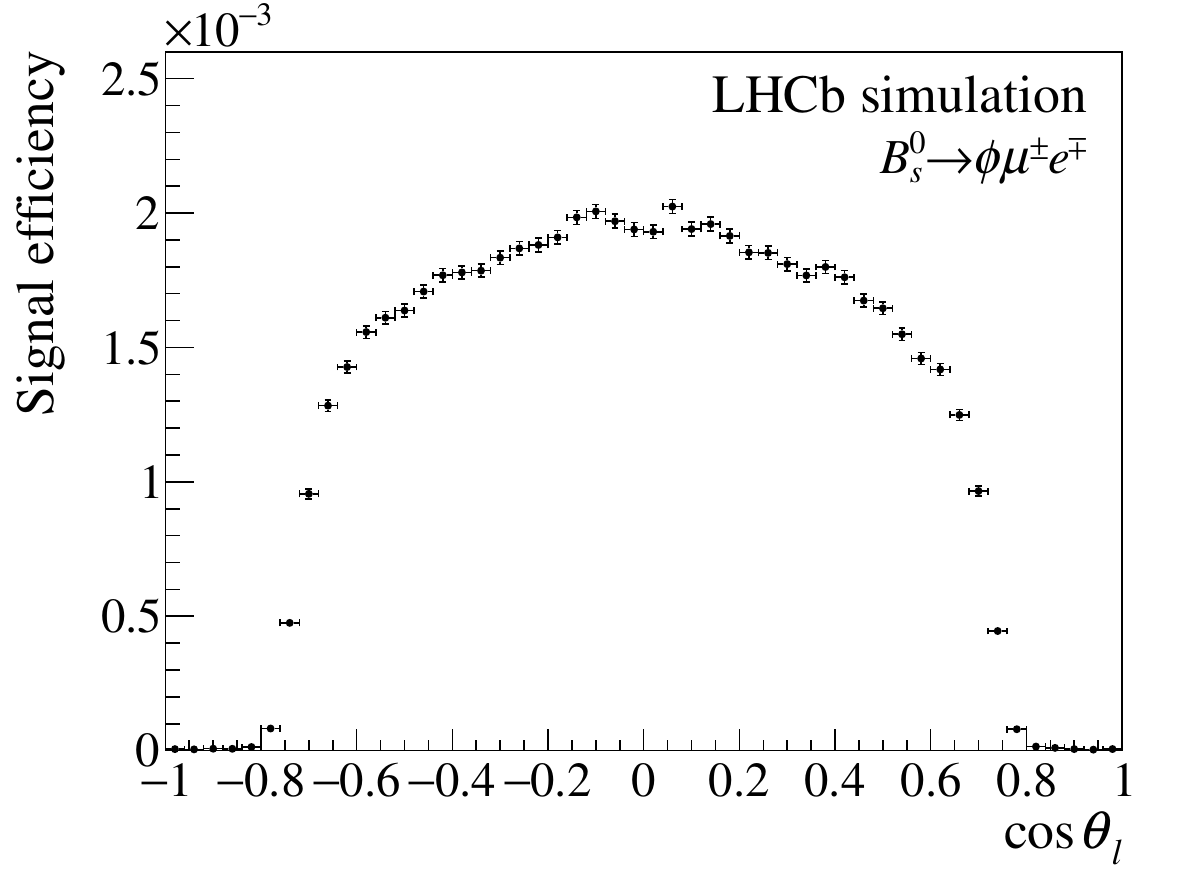}
  \includegraphics[width=0.4\textwidth]{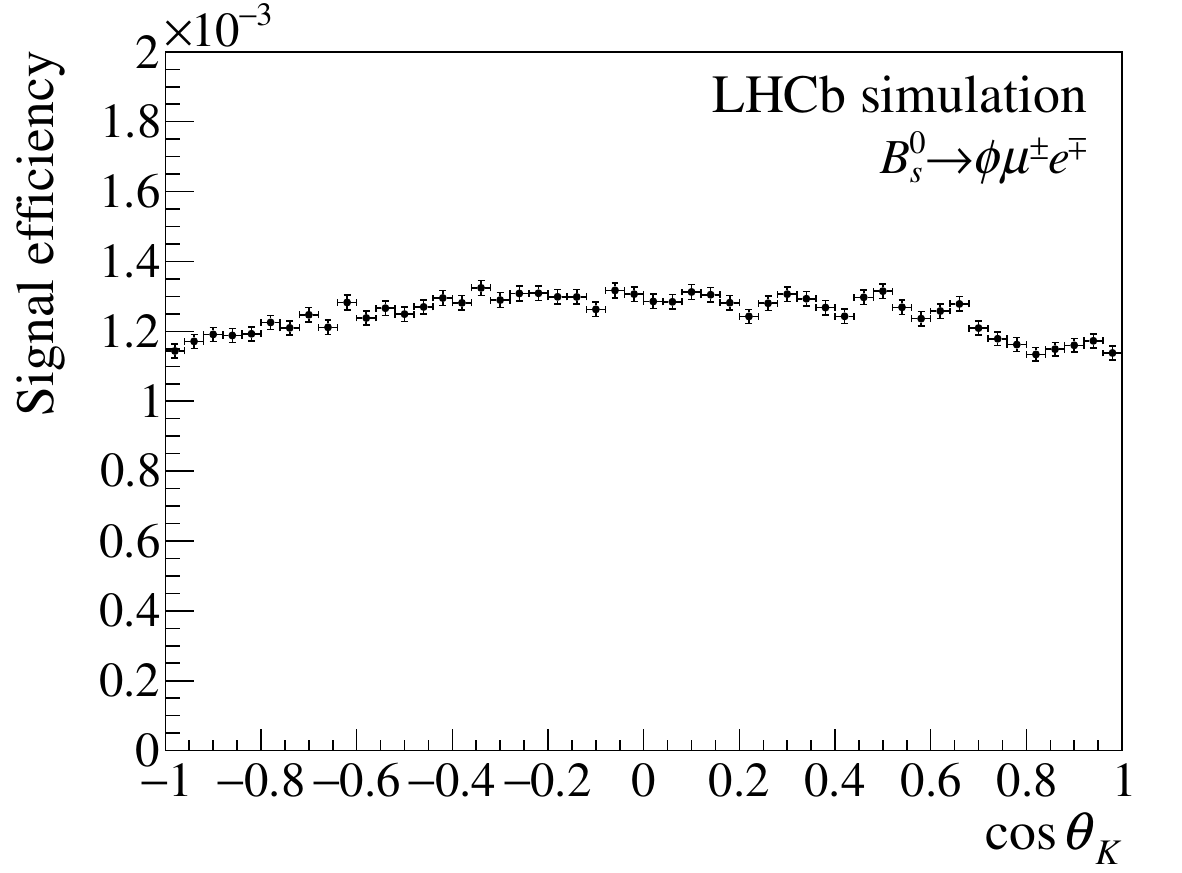}\\
  \caption{\small
    Reconstruction and selection efficiency, combined for the 2011 - 2018 samples, for (top) $\decay{\Bd}{\Kstarz\mu^\pm e^\mp}$ and (bottom) $\decay{\Bs}{\phi\mu^\pm e^\mp}$ signal decays depending on \qsq\ and the three decay angles $\ctl$, $\ctk$, and $\phi$.
    The selection efficiency drops in the \qsq\ region around the $\jpsi$ and $\psitwos$ masses as a result of the veto against misidentified backgrounds.
    \label{fig:efficiencies}}
\end{figure}
 
\newpage
\centerline
{\large\bf LHCb collaboration}
\begin
{flushleft}
\small
R.~Aaij$^{32}$\lhcborcid{0000-0003-0533-1952},
A.S.W.~Abdelmotteleb$^{50}$\lhcborcid{0000-0001-7905-0542},
C.~Abellan~Beteta$^{44}$,
F.~Abudin{\'e}n$^{50}$\lhcborcid{0000-0002-6737-3528},
T.~Ackernley$^{54}$\lhcborcid{0000-0002-5951-3498},
B.~Adeva$^{40}$\lhcborcid{0000-0001-9756-3712},
M.~Adinolfi$^{48}$\lhcborcid{0000-0002-1326-1264},
H.~Afsharnia$^{9}$,
C.~Agapopoulou$^{13}$\lhcborcid{0000-0002-2368-0147},
C.A.~Aidala$^{76}$\lhcborcid{0000-0001-9540-4988}
% ...

{\footnotesize \it

% ...
$^{9}$Universit{\'e} Clermont Auvergne, CNRS/IN2P3, LPC, Clermont-Ferrand, France\\
% ...
$^{13}$LPNHE, Sorbonne Universit{\'e}, Paris Diderot Sorbonne Paris Cit{\'e}, CNRS/IN2P3, Paris, France\\
% ...
$^{32}$Nikhef National Institute for Subatomic Physics, Amsterdam, Netherlands\\
% ...
$^{40}$Instituto Galego de F{\'\i}sica de Altas Enerx{\'\i}as (IGFAE), Universidade de Santiago de Compostela, Santiago de Compostela, Spain\\
% ...
$^{44}$Physik-Institut, Universit{\"a}t Z{\"u}rich, Z{\"u}rich, Switzerland\\
% ...
$^{48}$H.H. Wills Physics Laboratory, University of Bristol, Bristol, United Kingdom\\
% ...
$^{50}$Department of Physics, University of Warwick, Coventry, United Kingdom\\
% ...
$^{54}$Oliver Lodge Laboratory, University of Liverpool, Liverpool, United Kingdom\\
% ...
$^{62}$Syracuse University, Syracuse, NY, United States\\
% ...
$^{76}$University of Michigan, Ann Arbor, MI, United States, associated to $^{62}$\\
\bigskip
% ...
}
\end{flushleft}

\end{document}